\documentclass[cite,rivista]{cimento}
\usepackage{graphicx} 
\usepackage{subfloat}                              

\title{The Italian contributions to cosmic-ray physics \ETC\ from Bruno Rossi to the G-Stack. A new window into the inexhaustible wealth of nature}

\author{G.~Peruzzi\from{ins:x}\from{ins:e1}, \atque S.~Talas\from{ins:x}\from{ins:e2}}
\instlist{\inst{ins:x} Department of Physics - Padua University - Via Marzolo, 8 35131 Padua -Italy
\inst{ins:e1} peruzzi@pd.infn.it
\inst{ins:e2} sofia.talas@unipd.it}
\PACSes{\PACSit{01.65.+g}{History of science}\PACSit{10.00.00}{The physics of elementary particles and fields}\PACSit{29.00.00}{Experimental methods and instrumentation for elementary-particle and nuclear physics}}

\begin{document}

\maketitle

\begin{abstract}
From the late 1920s to the early 1950s, cosmic rays were the main instrument to investigate what we now call ``high-energy physics''. In approximately 25 years, an intense experimental and theoretical work brought particle physics from its childhood to its maturity. The data collected at that time played a crucial role in order to outline the quantum theories of  fundamental interactions - electromagnetic, weak and strong - and they were an excellent training field for many young scientists. The knowledge acquired on cosmic rays constituted the basis of the extraordinary progresses that were to be achieved in fundamental physics with the introduction of particles accelerators.

Many papers on this subject have already been published, both by historians and by the scientists who personally worked at that time. However, a complete historical reconstruction of those 25 years of cosmic-ray researches has not been written yet. In particular, an overall description of the researches carried out with the contributions of Italian physicists is still missing. The present paper tries to outline such a description, taking into account the fact that in those very years, for scientific and non scientific reasons, physics researches were getting organised within international collaborations. Without pretending to be exhaustive at all, we would like this paper to be a stimulus for further researches on the subject. 
\end{abstract}

\newpage

\tableofcontents

``The discoveries revealed by the observations here given are best 
explained by assuming that radiation of great penetrating power 
enters our atmosphere from outside and engenders ionization 
even in counters lying deep in the atmosphere. The intensity of 
this radiation appears to vary hourly. Since I found no diminution 
of this radiation for balloon flights during an eclipse or at night time we can hardly consider the sun as its source'' (cf. \cite{Hess1912}, English translation p. 672). With these words, Victor Franz Hess announced in 1912 the discovery 
of a mysterious radiation, which Robert Millikan was to call 
``cosmic rays'' in 1925. 

These new ``rays'', which were to play a crucial role in the further developments  of physics, burst onto the scene while physics was going through  a period of great effervescence. In 1905, Albert Einstein had postulated, on the basis of Planck's quantum hypothesis, that light, i.e. the electromagnetic radiation, can be treated in particular circumstances as consisting of individual quantum particles, which were to be called ``photons'' and, with some difficulty, the idea of the existence of this quantum 
of the electromagnetic field was finally accepted in the 1920s. As for 
matter, physicists still assumed, at the beginning of the 1920s, 
that it was only made of electrons and protons. However, the 
introduction of quantum mechanics led the physicists to revise 
their views on the elementary constituents of matter and on the 
number and the very nature of their fundamental interactions: 
new theoretic views and new experimental evidences seemed to 
hint at the necessity of radical changes.

It is worth giving here a short outline of the main features of these theoretical developments in order to understand some of the difficulties that were to be faced in those years within experimental researches, keeping in mind that the developments of theoretical schemas were of course closely correlated with the growth of new experimental results.  At the end of the 1920s only the electromagnetic interactions had been formulated in a quantum and relativistic theory, i.e. early quantum electrodynamics (early QED)\cite{Dirac1927, HeisenbergPauli1929, HeisenbergPauli1930, Fermi1932}. During the 1930s, admittedly with difficulties, Dirac's relativistic theory of the electron met early QED. In the same years the existence of the anti-electron was clarified as a consequence of Dirac's theory, the existence of the neutrino was postulated to interpret $\beta$ spectrum, and the discovery of the neutron led to the final interpretation of the nucleus as consisting of protons and neutrons.  Despite the problems of early QED (in particular, apparent internal inconsistency due to the emergence of infinities when the theory was used to calculate observable quantities), this theory on the one hand provided an important tool to interpret electromagnetic interactions (i.e. pair production, bremsstrahlung, Compton scattering), and on the other hand it was the reference model to develop quantum field theory within nuclear interactions (Fermi's theory of $\beta$-decay and Yukawa's theory of nuclear forces). However the theoretical knowledge was still rough with respect to the experimental data. This is why great difficulties were to be faced  in the different steps of the researches. An unbelievable effort was necessary to recompose in quite a coherent theoretical framework the different pieces of the puzzle of early particle physics.\footnote{For details, see \cite{Hoddeson1983, Pais1986, Hoddeson1989, Schweber1994, Rechenberg1996}.}

It is within this context that cosmic rays were studied 
and analysed. Until the end of the 1920s, they were simply regarded 
as $\gamma$ rays, i.e. high-energy photons, the most penetrating 
radiation known at that time. However, an experiment carried 
out by Walther Bothe and Werner Kohlh\"{o}rster in 1929 opened 
a new window on the nature of cosmic rays. The two German physicists 
showed that cosmic rays at ground level were charged particles, 
most of which (about 76\%) could penetrate through at least 4.1 
cm of gold. The observed penetrating power was thus enormously 
higher than that of the secondary particles produced by $\gamma$ 
rays; this led them to suppose that the primary rays had a corpuscular 
nature \cite{BK1929}. It is worth pointing out 
that the so called ``Geiger counters'' or ``Geiger-M\"{u}ller counters'' 
were crucial to carry out Bothe and Kohlh\"{o}rster experiment.\footnote{Rossi sticks to this point in\cite{Rossi1986}, pp. 48-9. Up to that moment, only ionization chambers had mainly been used. These are gaseous detectors based on the fact that a charged particle, when it traverses a gas, looses  energy as it ionizes the gas itself. An \textit{ionization chamber} mainly consists of two electrodes separated by a gaseous dielectric and kept at different potentials. It operates with a voltage which is large enough to collect all the electron-ion pairs, yet not so large as to produce any multiplication. As the voltage is further raised, the original free charges are multiplied because of  the interaction of the electrons which move through the gas toward the collecting electrode. Over a considerable range of voltage, the total number of collected electron-ion pairs is fairly proportional to the original ionization caused by the traversal of the charged particle. A detector operated in this region of voltage is called a \textit{proportional counter}: it has an advantage over ionization chambers in that the signals are much stronger (achievable gains on the order of $10^2$ to $10^3$). Finally, a further increase of the voltage leads to a region where very large multiplications are observed, and where the collected electron-ion pairs are independent of the original ionization. This is the region of \textit{Geiger-M\"uller counters}, which have the great advantage of a very large output pulse, so that their operation is simple and reliable (cf. \cite{Melissinos}, pp. 321-22).} These instruments, invented in 1908 by Hans Geiger and Ernest Rutherford \cite{RutherfordGeiger1908}, used the ionizing power of charged particles, and they detected and counted particles 
by electric methods.\footnote{A Geiger counter usually consists of 
a metal tube with a thin metal wire along its axis. The tube 
is filled with gas at low pressure (about a fraction of an atmosphere), 
and a potential difference (the so called ``operating voltage'' 
of some hundred of volts) is applied between the tube and the 
wire. This voltage is not high enough to produce an electric 
spark in the tube but, when a charged particle passes through 
the tube, it ionizes the gas, and an electric discharge is produced 
and registered by an electrometer.} These counters were largely 
diffused and, in the new and revised version developed in 1928 
by Geiger himself and Walther M\"{u}ller\cite{GeigerMueller1928a, GeigerMueller1928b}, they were fundamental for the study of cosmic rays. 

The experiment of Bothe and Kohlh\"{o}rster was the starting point 
of a lively debate about the nature of cosmic rays and opened 
a controversy: were cosmic rays particles or very high energy 
electromagnetic waves, i.e. $\gamma$ rays as Millikan supported? 
Once again, physicists had to face the wave-particle dilemma 
in respect of new physical phenomena.\footnote{An analogous wave-particle debate had been carried out about the nature of light from the eighteenth 
to the early nineteenth century, and the nature of cathode rays was 
debated in a similar way in the second half of the nineteenth 
century.} The Bothe and Kohlh\"{o}rster experiment was a challenge 
to the dominant interpretative model, and several physicists 
started working on cosmic rays. A young Venetian scientist, Bruno 
Rossi, felt fascinated by the new challenge too, and he soon 
became an outstanding pioneer in cosmic rays researches. As he 
wrote years later, he felt ``a subconscious feeling for the inexhaustible 
wealth of nature, a wealth that goes far beyond the imagination 
of man. That feeling [\dots] was the reason why, as a young man, 
I went into the field of cosmic rays. In any case, whenever technical 
progress opened a new window into the surrounding world, I felt 
the urge to look through this window, hoping to see something 
unexpected'' (cited by George W. Clark in \cite{Clark1998}, p. 311). 

\section{The beginnings of cosmic-ray researches in Italy: from 
the late 1920s to 1945}

\subsection{Bruno Rossi's researches from 1929 to 1938}

Born in Venice in 1905, Bruno Benedetto Rossi studied in Padua 
and Bologna. He started his academic career in 1928 at the University 
of Florence, where he worked with a group of physicists he particularly 
appreciated for their scientific and human qualities (fig. 1). The group 
was composed of Gilberto Bernardini, who was to become one of 
the prominent scientists of Italian post-war physics, Giulio 
Racah, Daria Bocciarelli, Lorenzo Emo, Guglielmo Righini, Beatrice 
Crin\`{o} and Giuseppe (Beppo) Occhialini. The latter, who was to carry out fundamental studies on cosmic rays, always considered himself as Rossi's 
pupil, though he was only two years younger. 

\begin{figure}

\centering

\includegraphics[width=\textwidth]{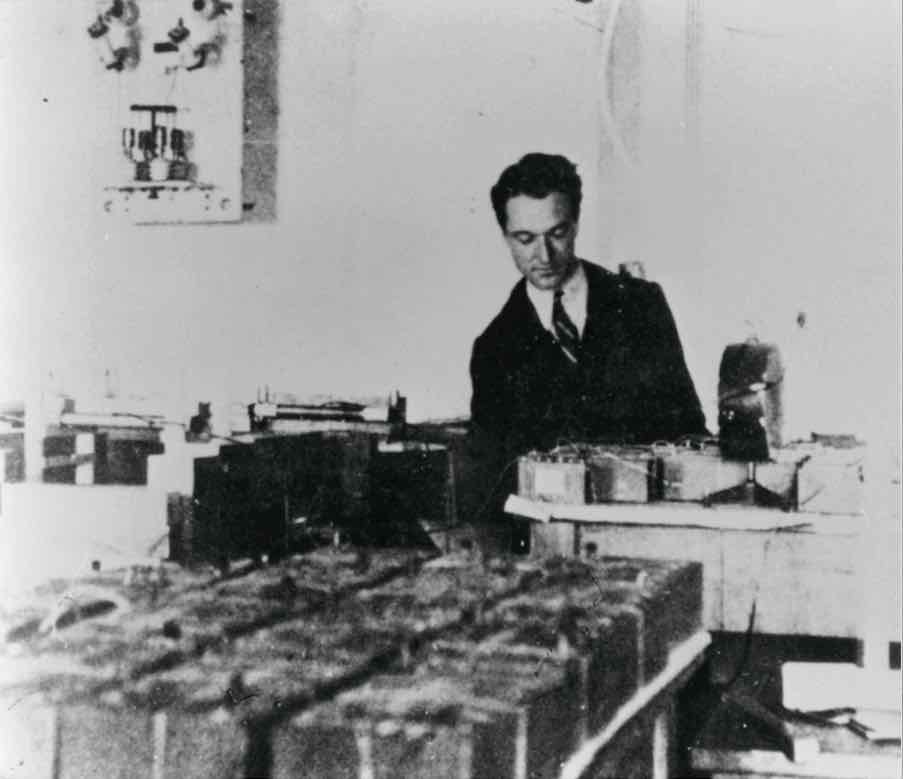}

\caption{Bruno Rossi working on cosmic rays at the University of Florence (reproduced from \cite{Rossi1987}).}\label{fig:1}

\end{figure}

Within a few weeks after Bothe and Kohlh\"{o}rster experiment, 
Rossi invented and published the design of an electronic coincidence 
circuit composed of triodes and Geiger-M\"{u}ller counters, which 
recorded the simultaneous occurrence (coincidence) of three or 
more electrical pulses arriving from the different counters (cf. \cite{Rossi1930a} and fig. 2). Rossi's circuit marked the beginning of the use 
of electronic devices in nuclear and particle experimental physics and, shortly later, it became one of the basic logic circuits of modern electronic computers (in fact, it is essentially an AND logic circuit). 

\begin{figure}

\centering

\includegraphics[width=\textwidth]{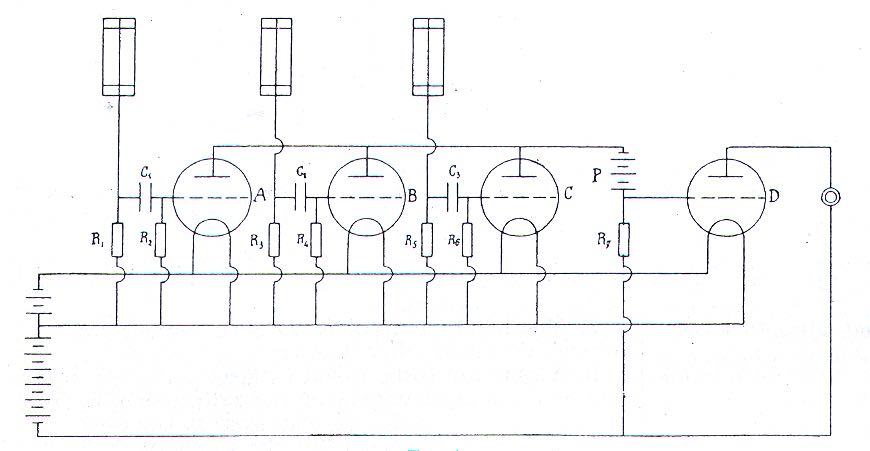}

\caption{Rossi's coincidence circuit. ``$R_1 = R_3 = R_5 = 5\times 10^9 \mbox{ohms}$, $R_2 = R_4 = R_6 = R_7 = 8\times 10^6 \mbox{ohms}$, $C_1 = C_2 = C_3 = 10^{-4} \mu F$. The positive electrodes of the three counters [\dots] are electrostatically coupled to the grids of the three valves $A$, $B$, $C$. In normal conditions these grids have a zero potential; whenever a discharge occurs they become negative, thus interrupting the current flow. As the resistance $R_7$ is very great compared with the internal resistances of the valves $A$, $B$, $C$, their anodes are at a potential near to zero. The grid of the valve $D$ (for the introduction of the auxiliary battery $P$) is at a slight negative potential. This potential varies very little when only one or two counter tubes are working, while it undergoes a sudden rise when, for the simultaneous working of the three counter tubes, the current is interrupted in all the three valves. The consequent variation of the anode current [\dots] is acoustically detected by a telephone'' (the diagram is reproduced from \cite{Rossi1930a}).}\label{fig:2}

\end{figure}

With this new device, which was to be widely used in cosmic-ray 
physics, Rossi carried out several experiments. He discovered 
for instance that the interaction of cosmic rays with matter 
produced groups of particles -- later called \textit{showers} -- with 
an unexpected frequency.\footnote{This result seemed so astonishing 
that it was accepted for publication only by Werner Heisenberg's 
good offices (see \cite{Rossi1932a}).} To perform this experiment, inspired by the tracks observed by Dmitry V. Skobelzyn in a cloud chamber\cite{Skobelzyn1929}, Rossi had placed three Geiger counters in a triangular 
configuration, so that they could not all three be discharged 
by a single particle travelling in a straight line, and he had 
enclosed the setup in a lead box. Through another experiment, 
Rossi demonstrated that cosmic-ray particles could traverse more 
than 1 meter of lead\cite{Rossi1932b}. This reinforced his supporting the particle thesis for cosmic rays -- i.e. the thesis according to which the primary 
cosmic radiation was a flux of charged particles --,\footnote{The ``primary 
cosmic radiation'' is the radiation that arrives on the surface 
of Earth's atmosphere, while the so called ``secondary radiation'' 
is produced by the interactions between the primary radiation 
and the constituents of atmosphere.} against the wave thesis, 
which regarded the primary radiation as consisting of $\gamma$ 
rays. Let us point out that Arthur Compton became at about that 
time one of the most strenuous supporters of the particle thesis, 
while Robert Millikan fervently went on defending the wave thesis (fig. 3).

\begin{figure}
\centering

\includegraphics[scale=0.43]{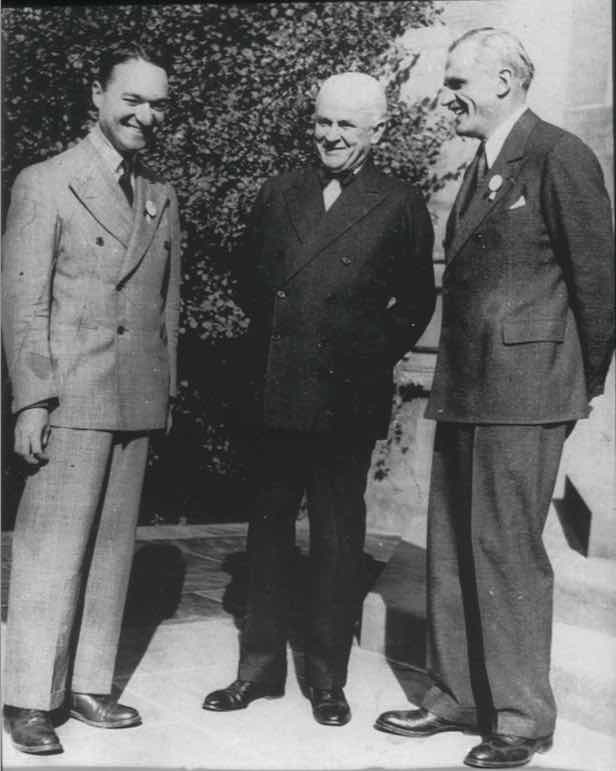}

\caption{Bruno Rossi with Robert Millikan and Arthur Compton in 1931 (reproduced from \cite{Rossi1987}).}\label{fig:3}

\end{figure}

A way to clear the controversy about the nature of primary cosmic rays was proposed by Rossi in the early 1930s. In fact, the physicists already knew that, if the primary cosmic rays were particles, they were to be deviated by the earth magnetic field, so that a decrease of the intensity of cosmic rays was expected when approaching the magnetic equator. The existence of this {\it latitude effect} was under examination in those very years. As for Rossi, he suggested that, if the cosmic primary radiation was constituted of charged particles, an azimuthal effect -- the so called {\it east-west effect} -- was to be expected as well, i.e. an asymmetry in the intensity of the cosmic radiation with respect to the magnetic meridian: a larger intensity of the radiation was expected from east in case of prevailing negative charged particles, or from west in case of prevailing positive charged particles. According to Rossi's theory, the east-west asymmetry was greater at lower geomagnetic latitudes \cite{Rossi1930b,Rossi1931}.\footnote{An analysis of the motion of charged particles in the magnetic field of the Earth had been made by Carl St\"ormer in connection with the theory of the aurora borealis (a summary of St\"ormer's early results from 1906 onwards is given in \cite{Stoermer1930}). Rossi's theory was probably the first application of St\"ormer's results to study the nature of cosmic rays. A further extension was performed in 1933 by Georges Lema\^{\i}tre and Manuel Sandoval Vallarta\cite{LemaitreVallarta1933} (see \cite{Vallarta1938} for a summary of the further developments), who did not seem to know about Rossi's published works.} Rossi himself immediately 
tried to validate his theory at Florence in 1931, but unsuccessfully \cite{Rossi1931}. In a subsequent paper written with Enrico Fermi \cite{FermiRossi1933} (fig. 4), he explained that the negative result of the experiment was due to the atmospheric absorption: it was therefore of the 
utmost importance to carry out observations at sufficiently low 
latitudes -- where the effect was expected to be greater -- and 
at great altitudes with respect to the sea level. Rossi thus 
conceived the project of an expedition near the magnetic equator.

\begin{figure}
\centering

\includegraphics[width=\textwidth]{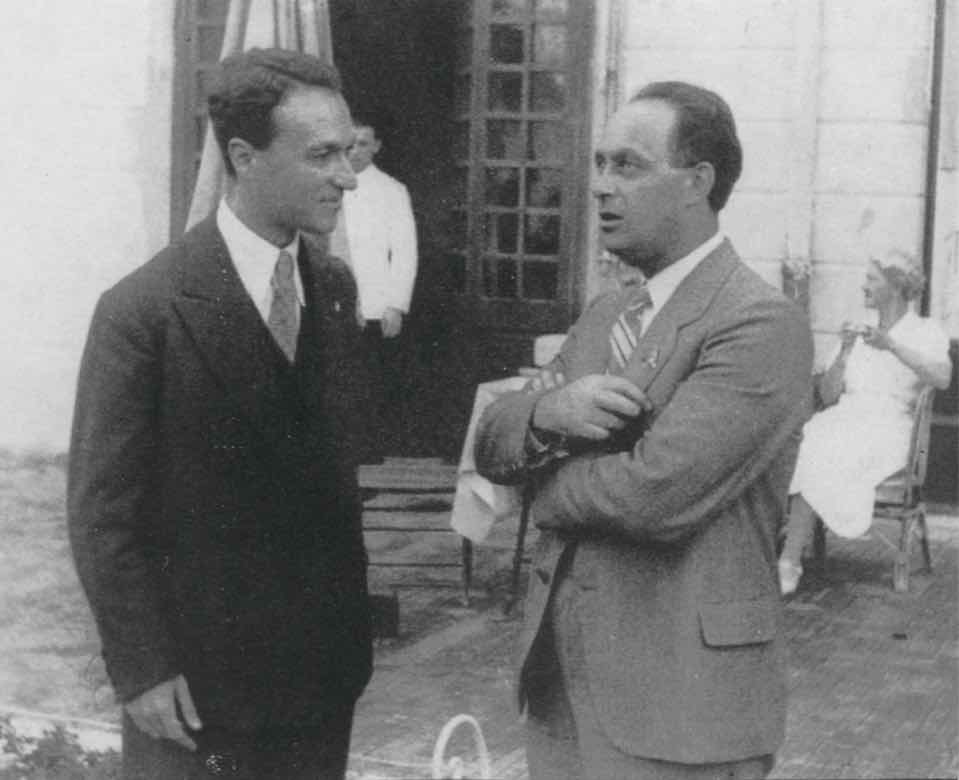}

\caption{Bruno Rossi and Enrico Fermi in the 1930s (reproduced from \cite{Rossi1987}).}\label{fig:4}

\end{figure}

Meanwhile, in 1932, he was appointed to the chair of Experimental 
Physics at the University of Padua, where he introduced his experimental 
researches on cosmic rays,\footnote{These were a completely new field 
of research in Padua, as Giuseppe Vicentini, the holder of the 
chair of Experimental Physics up to 1931, had mainly focused his 
work on X-rays and seismography.} and in 1933, he succeeded in 
organising, with the financial support of the Italian National 
Research Council (CNR), an expedition in Eritrea intended to study 
the \textit{latitude effect} and the \textit{east-west effect}.\footnote{Concerning the expedition in Eritrea, see the various articles of Rossi and 
Sergio De Benedetti in \textit{La Ricerca Scientifica}, anno IV (1933) and anno V (1934). } As for the latitude effect, experiments had already been widely carried out from the late 20s and most of the results had been negative, with the exception of the results published by Jacob Clay from 1927 (see \cite{Clay1930}).\footnote{The results obtained by Clay in 1927-29 are now regarded as controversial and unreliable, as pointed out by Ad Maas in his talk ``Machinations, Manipulations, and Cosmic-Ray Measurements. Jacob Clay and the Discovery of the Latitude Effect'' given at the {\it XXVI Symposium of the Scientific Instrument Commission of the International Union for the History and Philosophy of Science} held in Cambridge, Massachussets, on 6-11 September 2007.} The situation had thus remained quite confused until the worldwide coordinated study organised by Compton from the summer of 1930 (see \cite{Compton1933,Johnson1938} for a complete bibliography on the subject). Compton's results, published from 1932 onwards, showed definite differences in the intensity of the cosmic rays at different latitudes (see \cite{Compton1932a,Compton1932b,Compton1933}). The Italian expedition, guided by Bruno Rossi himself, confirmed the gradual decrease of the intensity of the radiation when approaching the magnetic equator during the voyage from Spalato to Massaua. As for the east-west effect, the Italian scientist studied it at 
Asmara, in Eritrea, near the magnetic equator (fig. 5). The measurements were performed at the altitude of 2370 metres above the sea level and at the geomagnetic latitude of 11$^\circ$30' north. The 
experiments showed that, for a given zenithal angle, the intensity 
of the cosmic radiation coming from west was quite greater than 
that coming from east: this seemed to prove that cosmic rays mainly consisted of positive charged particles \cite{Rossi1934a}. Unfortunately, Rossi's moving from Florence to Padua, and the difficulties in obtaining the 
financial support from the government (quite a constant problem 
in the history of our country) had delayed the expedition, and 
such a delay cost Rossi, with much sadness, the priority of the 
discovery. Actually, a short time before the beginning of Italian 
experiments, other groups of scientist, in particular Thomas 
Johnson \cite{Johnson1933}, and the group led by Compton \cite{AlvarezCompton1933}, had obtained similar results. Nevertheless, according 
to Rossi, the effect detected by the other groups was much smaller 
than the one measured by the Italian, so that it left room to 
Compton's hypothesis, which suggested that only a small fraction 
of cosmic rays was composed of positive charged particles. Rossi's 
experiment was thus important to define the actual nature of the primary 
radiation (fig. 6, 7).\footnote{It is worth noting that, in spite of Rossi's results, the question of the sign of the charge of the primary particles remained to be answered with certainty for several years. New successful measurements of the east-west effect, carried out in particular by T.H. Johnson and collaborators\cite{Johnson1938}, confirmed the deduction that most, if not all, of the primary radiation was positive. However, such a conclusion could not be regarded as certain because the observed radiation was mostly secondary or later-stage radiation, and it was not known to what extent this later radiation preserved the direction of the primaries, or whether the degree of multiplicity production of secondaries was the same for both positive and negative primaries (on this point see for instance \cite{Montgomery1943}, p. 8 and p. 11, and see also the end of section 1.2 of the present paper).} 
\begin{figure}
\centering

\includegraphics[width=\textwidth]{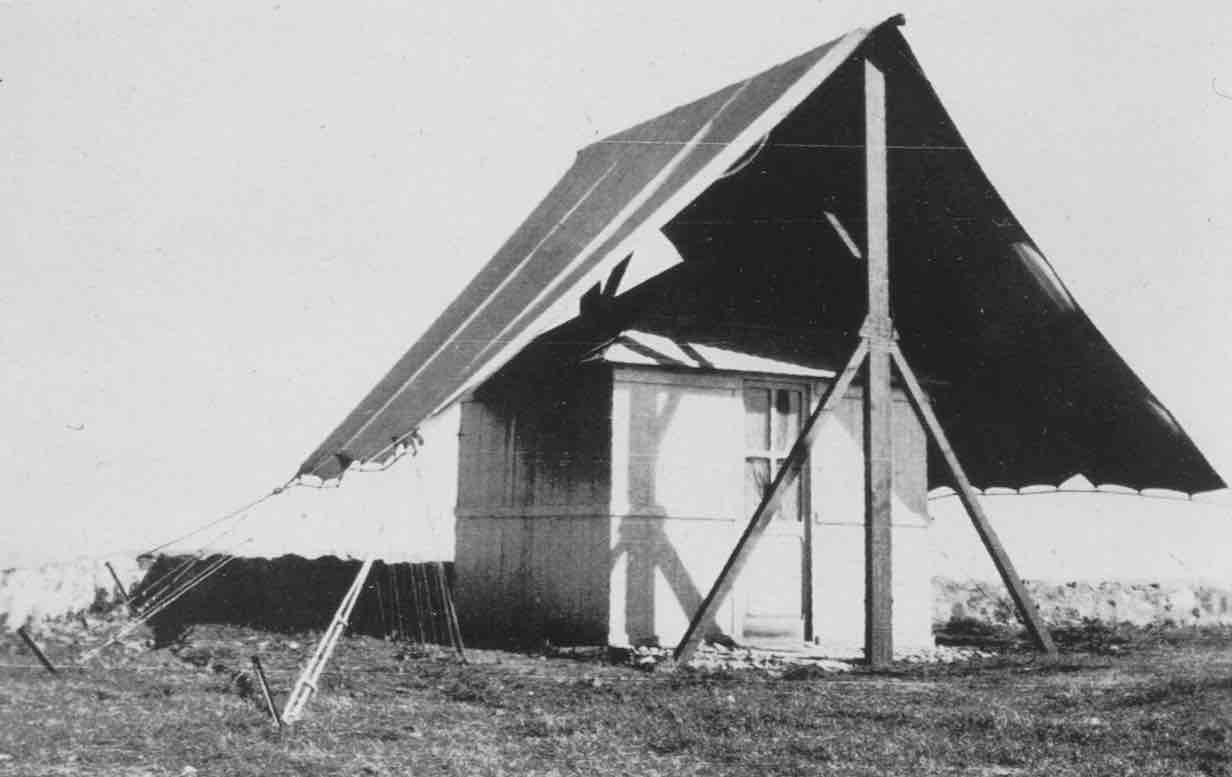}

\caption{The light wood hut in Asmara, Eritrea, that housed Rossi's experiments on the angular distribution of cosmic ray intensity (east-west effect), 1933 (reproduced from \cite{Rossi1934b}, p. 581).}\label{fig:5}

\end{figure}
\begin{figure}
\centering

\includegraphics[scale=0.5]{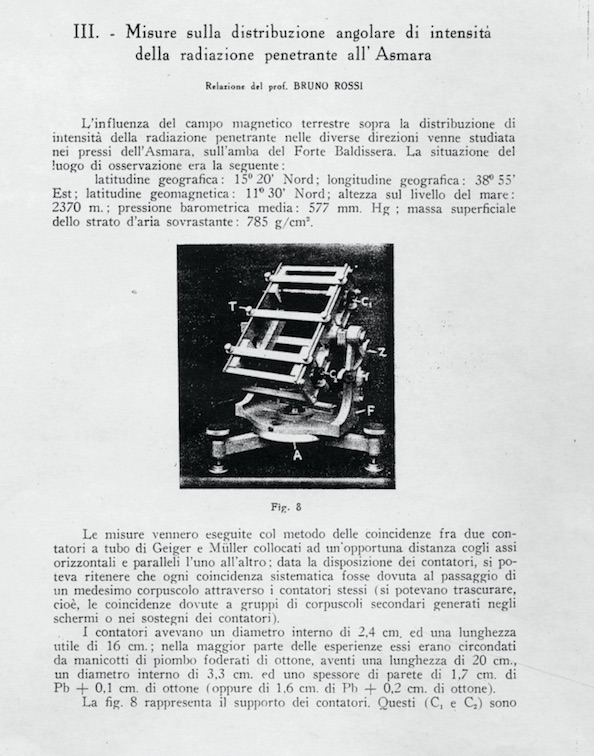}

\caption{A page reproduced from one of Rossi's papers, published in 1934, relating the experiments he carried out in Eritrea on the east-west effect (\cite{Rossi1934b}, p. 579). The instrument described in this page is the alt-azimuthal support for two Geiger-M\"uller counters used for the measurements.}\label{fig:6}

\end{figure}
\begin{figure}
\centering

\includegraphics[width=\textwidth]{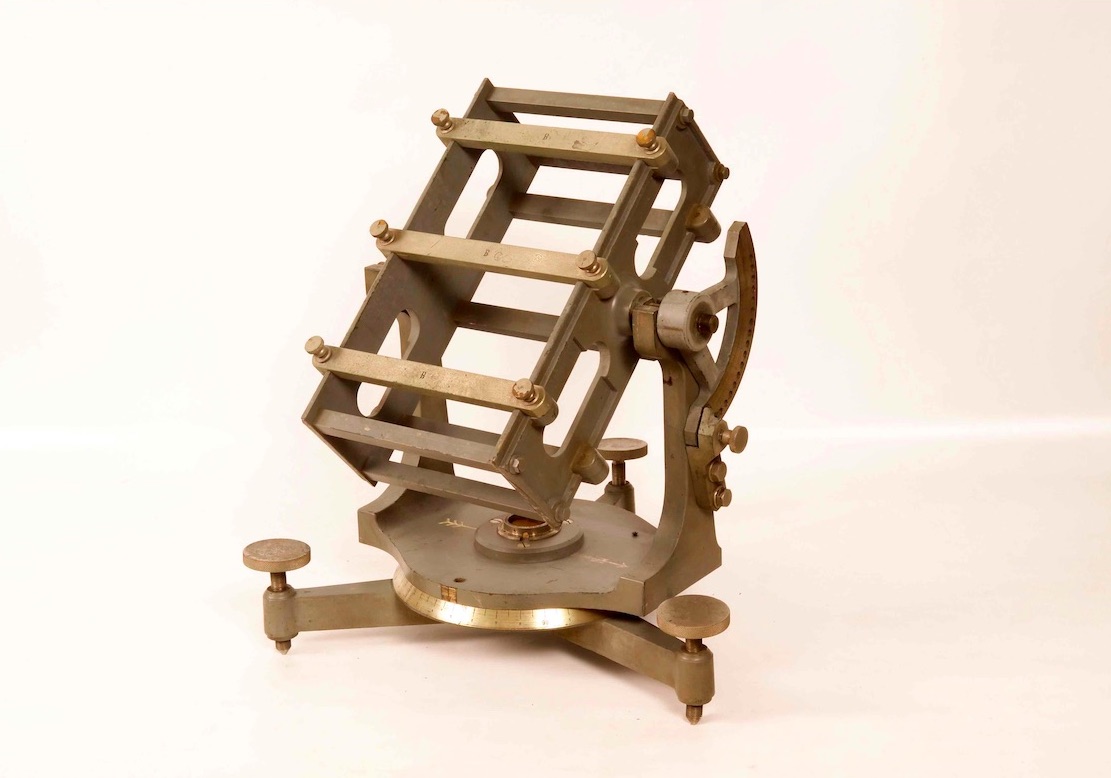}

\caption{The alt-azimuthal support for Geiger-M\"uller counters, 1933. Two counters were mounted in parallel one upon the other and the system they formed could be moved vertically and horizontally. Two graduated scales gave the zenithal and the azimuthal angle of the two counters system. One could thus study the intensity of cosmic rays coming from different directions (Museum of the History of Physics, University of Padua).}\label{fig:7}

\end{figure}

\begin{figure}
\centering

\includegraphics[scale=0.5]{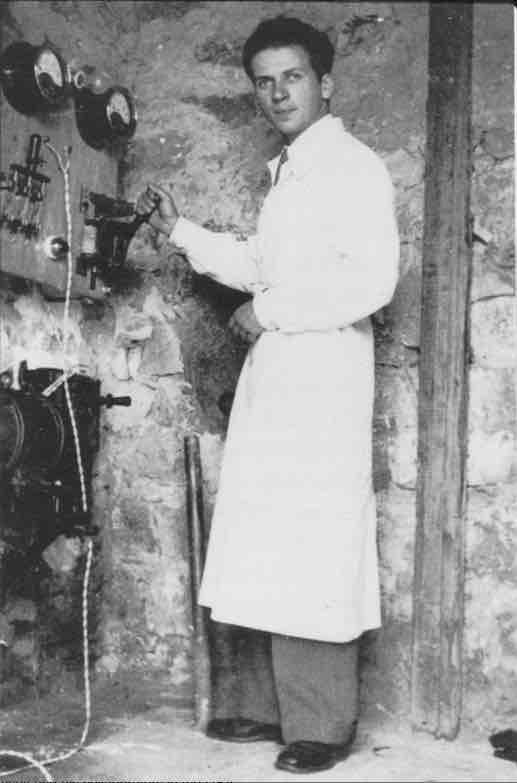}

\caption{Sergio De Benedetti inside the Asmara hut, 1933 (reproduced from \cite{Rossi1987}).}\label{fig:8}

\end{figure}

It is worth pointing out that the Eritrea expedition also led 
to the first conjecture of the existence of extensive showers 
in cosmic radiation. In fact, as the measurements had to take 
account of the chance coincidences, Rossi determined the frequency 
of these coincidences by using two Geiger-M\"{u}ller counters far 
from one another. Surprisingly, Rossi and his young collaborator, 
Sergio De Benedetti (fig. 8), observed that the frequency of the chance 
coincidences was greater than what was predicted on the basis 
of the resolving time of the coincidence circuit. This gave a 
hint that not all coincidences were actually chance coincidences. 
Moreover, the frequency of the coincidences registered by means 
of three counters appeared to be too great compared to the 
frequency registered by means of two counters. Rossi himself 
proposed that very extended showers occasionally arrived upon 
the instruments. Rossi's hypothesis was confirmed at the end 
of the 1930s by the work of Pierre Auger and Roland Maze, who 
analysed for the first time the structure and dimension of extensive air 
showers \cite{AugerMaze1939}. It is worth mentioning, 
stopping just for a moment the chronological narration, that 
the very extended air showers were to be fundamental in the study 
-- carried out by Rossi himself and his collaborators in the USA 
by the end of the 1940s -- of the highest energy part of the cosmic 
radiation spectrum: actually, the measurements on extensive showers were precious in giving indirectly quantitative details on the primary cosmic-ray particles of very high energy. As Rossi wrote several years later ``air showers experiments [\dots ] provide the only available means of detecting primary cosmic-ray particles of the highest energies and of determining their energy 
spectrum and their arrival directions'' (cf. \cite{Rossi1982}, p. 82). 
Nowadays, as accelerator physics has to face enormous costs, 
there is a renewed interest in cosmic ray physics and, in particular, 
in extensive showers.

By the early 1930s, several properties of cosmic rays had thus 
been discovered and scientists were growing more and more interested 
in the physical nature of cosmic rays. In those very years, the 
use of cloud chambers within cosmic rays researches was to bring 
fundamental discoveries in the field, as these devices could 
not only detect but also photograph the particles' tracks. The 
cloud chamber had been invented in 1897 by Charles T.R. Wilson 
to study the physics of meteorological phenomena related to the 
formation of fog. In Wilson's device, a quick cooling down of 
the vapour was achieved through a fast expansion. In particular, 
Wilson analysed what happened when very pure vapour was brought 
to temperatures below the temperature of condensation without 
condensing. The vapour was then in an unstable state, the 
so called ``supersaturated'' state, where any impurity constituted 
a nucleus of condensation. Wilson observed that charged particles 
could act as condensation nuclei and, in 1911, he proposed to 
use this phenomenon to visualize the track of particles which 
passed through a gas and ionized it \cite{Wilson1911}. The track of a particle 
was thus seen as a thick row of drops. For this device, Wilson 
was awarded the Nobel Prize for Physics in 1927.\footnote{In 1952, George D. Rochester and John G. Wilson published a volume containing both particular photographs of historic importance, and photographs illustrating typical examples of the principal phenomena discovered with cloud chambers in cosmic ray physics up to that time \cite{RochesterWilson1952}.} 

A first, fundamental result on cosmic rays was obtained with 
a cloud chamber in 1932. A new particle, never observed before, 
was photographed by Carl Anderson in a cloud chamber immersed 
in a magnetic field \cite{Anderson1932}. The analysis of the track  showed that it was a positive particle with a mass similar to the mass of the electron. The new particle was later to be called \textit{anti-electron} or \textit{positron}. Meanwhile, at the end of 1931, Giuseppe Occhialini had moved to the Cavendish Laboratory in Cambridge, where he had started working with Patrick M. S. Blackett, who was a specialist of cloud chambers researches.  From Florence, Occhialini had brought in England his knowledge of Rossi's coincidence circuit and, with Blackett, he devised a cloud chamber with two 
Geiger-M\"uller tubes, placed one above and the other below the chamber 
itself. As an electrically charged particle passed through both Geiger-M\"uller 
tubes, the impulses from the two  counters coincided and the chamber was then brought into function:  the chamber was thus \textit{counter-controlled}. As Blackett and Occhialini wrote, they had ``developed a method by which the high speed particles associated with penetrating radiation can be made to take their own cloud photographs'' (cf. \cite{BlackettOcchialini1933}, p. 699; see also \cite{BlackettOcchialini1932}). Thanks to this new 
device, which was to prove extraordinarily useful for all further 
researches on these topics, Blackett and Occhialini found tracks 
on 80 per cent of the photographs (i.e. over 500 photographs), 
and they confirmed in the spring 1933 Anderson's discovery that 
``particles must exist with positive charge but with a mass comparable 
with that of an electron rather with that of a proton'' (cf. \cite{BlackettOcchialini1933}, p. 705). 
Moreover, as for the showers, according to the two scientists, 
``the main beam of downward moving particles consists chiefly 
of positive and negative electrons'' (cf. \cite{BlackettOcchialini1933}, p. 708). Therefore, as the ``positive electrons'' observed in the showers ``can only 
have a limited life as free particles since they do not appear 
to be associated with matter under normal conditions'', Blackett 
and Occhialini concluded that it was ``likely that they disappear 
by reacting with a negative electron to form two or more quanta. 
This latter mechanism is given immediately by Dirac's theory 
of electrons'' (cf. \cite{BlackettOcchialini1933}, p. 714).\footnote{Dirac's relativistic electron theory was proposed in 1928 \cite{Dirac1928}. As noted by Blackett and Occhialini in \cite{BlackettOcchialini1933} p. 714, in Dirac's theory all but a few of the quantum states of negative kinetic energy are taken to be filled with negative electrons. The few states which are unoccupied behave like ordinary particles with a positive charge. Dirac originally \cite{Dirac1930} wished to identify these ``holes'' with protons, but this had to be abandoned when it was found that the holes necessarily had the same mass as negative electrons \cite{Dirac1931}. ``It will be a task of immediate importance -- Blackett and Occhialini concluded -- to determine experimentally the mass of the positive electrons'' (cf. \cite{BlackettOcchialini1933} p. 714). It is worth pointing out that Anderson did not make any reference to Dirac's theory in his papers on the ``positive electron''.} As a matter of fact, they discussed this idea with Dirac himself. Though their measurements of the mass of the positron were not so accurate to assess with certainty its equality with that of the electron, they observed that ``no difference between the ionization from tracks of negative and positive electrons'' of the same curvature ``has been detected so that provisionally their masses may be taken as equal'' (cf. \cite{BlackettOcchialini1933}, p. 714). Moreover, Blacket and Occhialini learnt from Dirac the result of calculation he had made of the actual probability of the annihilation process of a positive and negative electron giving the values of the mean free path for annihilation, of the range, and of the mean life of positive electrons. They concluded that to test Dirac's prediction further detailed investigations were needed, but ``there appears to be no evidence as yet against its validity, and in its favour is the fact that it predicts a time of life for positive electron that is long enough for it to be observed in the cloud chamber but short enough to explain why it had not been discovered by other methods'' (cf. \cite{BlackettOcchialini1933}, p. 716).

In the meantime, Rossi had improved the experimental apparatus 
he had used to study cosmic rays showers \cite{Rossi1932a}, and 
he had measured the rate of triple coincidences as a function 
of the thickness of the lead above the counters  (fig. 9a). The new results, 
which came to be known as ``the Rossi curve'', were published in 
1933 \cite{Rossi1933a, Rossi1933b} (fig. 9b).  
\begin{subfigures}
\begin{figure}
  \includegraphics[width=\textwidth]{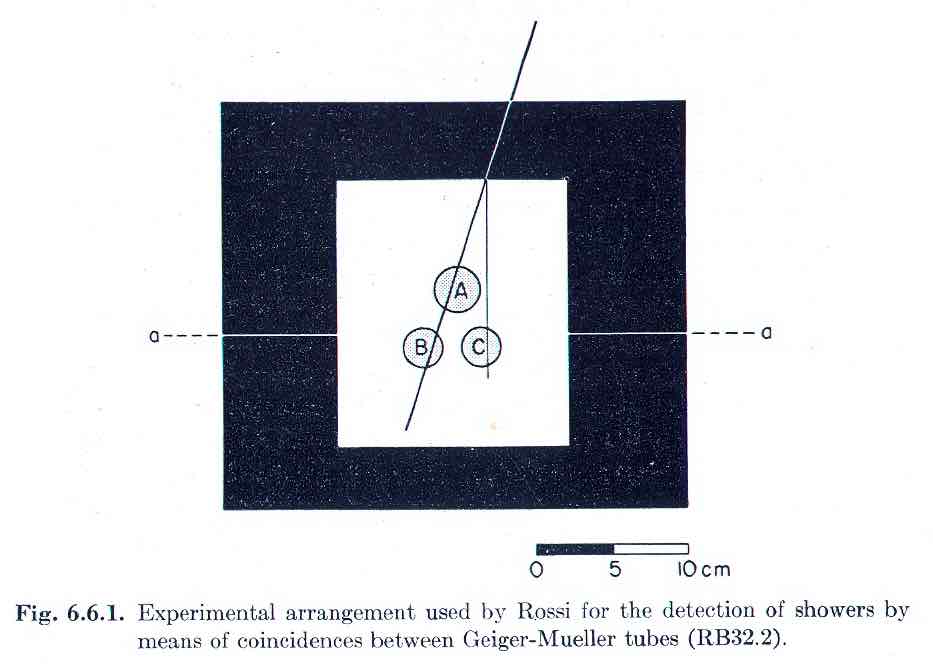}
  \caption{\label{fig:9a}Rossi's triangular arrangement of three Geiger-M\"uller counters reproduced from \cite{Rossi1952} which cites as a reference RB32.2 = \cite{Rossi1932a}.} 
\end{figure}
\begin{figure}
 \includegraphics[width=\textwidth]{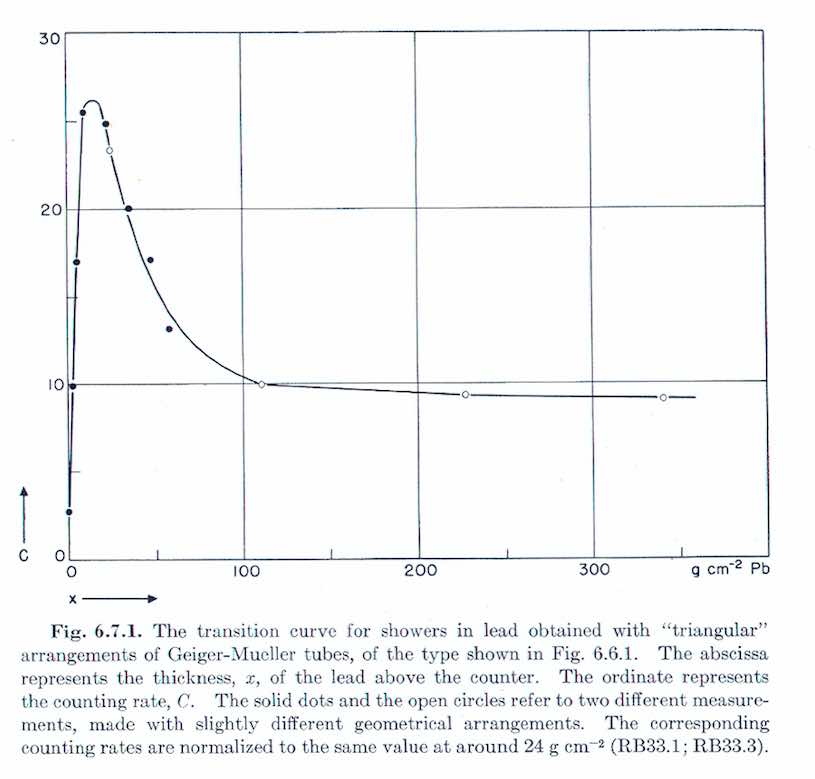}
  \caption{\label{fig:9b}The Rossi curve, also called ``transition curve'', reproduced from \cite{Rossi1952} which cites as references RB33.1 = \cite{Rossi1933a} and RB33.3 = \cite{Rossi1933c}.} 
\end{figure}
\end{subfigures}
They showed 
that, as the thickness of the lead increases, the ``number of 
showers emerging [\dots ] increases at first, reaches a maximum 
[\dots ] and then decreases very rapidly'' (cf. \cite{Rossi1933b}, p. 173). This 
led Rossi to conclude that the local cosmic rays consist of two 
components, the penetrating one, also called \textit{hard} component, 
which only occasionally gave rise to a shower, and a \textit{soft} component, 
of unknown nature, which prolifically generated particle showers 
and was rapidly attenuated in lead. As the frequency 
of the showers decreased very slowly when the thickness of the 
lead was further increased, Rossi regarded the shower-producing 
rays as a secondary radiation produced from the penetrating rays, 
which were regarded as the primary radiation. However, further 
experiments carried out in Padua in collaboration with several 
young scientists led Rossi to change some his views.\footnote{See 
the various papers by Giulia Alocco, Angelo Drigo, Bruno Rossi, Giovanni Bottecchia, and Sergio De Benedetti in \textit{La Ricerca Scientifica}, \textbf{V.1 and V.2} (1934) -- in particular see Rossi and De Benedetti \cite{RossiDeBenedetti1934} -- and the paper by B. Rossi in \textit{Nuovo Cimento} \cite{Rossi1934c}.} In particular, with De Benedetti, 
Rossi measured the intensity of both kind of radiations at increasing 
altitudes,\footnote{The measurements were carried out in Padua (40m above sea level),  at Passo della Mendola (1350m above sea level), and at Passo dello Stelvio (2756m above sea level).} and he observed that the intensity of the soft radiation increased much more rapidly than the intensity of the penetrating hard radiation, so that the two radiations seemed to be regarded 
both as primary radiations, independent one from another. 

A few years later, in October 1937, a Conference was held in 
Bologna on the occasion of the second centenary of Luigi Galvani's 
birth, and several of the most important physicists of those 
years were invited to give a talk -- let's mention for instance 
Niels Bohr, John D. Cockcroft, Peter Debye, Enrico Fermi, Wolfgang Pauli, Werner Karl Heisenberg, Erwin Schr\"odinger, 
Arnold Sommerfeld. As for Rossi, he presented a general survey on cosmic 
ray researches, and he spoke again about ``due tipi di corpuscoli 
elettrizzati: i corpuscoli duri, dotati di un elevatissimo potere 
di penetrazione, che generano nella materia un numero relativamente 
piccolo di sciami [\dots ]; i corpuscoli molli, che generano un 
gran numero di sciami, che hanno un potere di penetrazione assai 
minore[\dots ]'' (cf. \cite{Rossi1938a}, p. 60-1).\footnote{``two kinds of electrified corpuscles: the hard corpuscles, which have a very high penetrating power and generate in matter a relatively small number of showers [\dots]; the soft corpuscles, which generate in matter a large number of showers and have a much lower penetrating power [\dots]''.}  Little had 
been known about the nature of the soft radiation until the introduction 
of the theory proposed by Hans Albrecht Bethe and Walter Heitler 
in 1934, which was further developed by Homi Jehangir Bhabha 
and Heitler, and then by J. Franklin Carlson and J. Robert Oppenheimer 
in 1937 \cite{BetheHeitler1934,BhabhaHeitler1937,CarlsonOppenheimer1937}. Rossi said at the 1937 Conference that ``una teoria recentemente sviluppata da Bhabha e Heitler e da Carlson e Oppenheimer, sulla base dell'elettrodinamica quantistica, conduce a prevedere per gli elettroni di grande energia un comportamento analogo a quello che sperimentalmente si trova per i raggi molli. Secondo questa teoria, la forma principale di interazione fra gli elettroni e la materia consisterebbe nella 
emissione di una radiazione $\gamma$ ``di frenamento''; i quanti $\gamma$, a 
loro volta, subirebbero un processo di materializzazione dando 
origine ad una coppia di elettroni, i quali emetterebbero di 
nuovo raggi $\gamma$, e cos\`{\i} via. Gli sciami nascerebbero appunto 
dalla successione alternata di questi due processi di radiazione 
di frenamento e di materializzazione di energia; la radiazione 
molle locale sarebbe quindi composta di elettroni dei due segni 
e di raggi $\gamma$'' (cf. \cite{Rossi1938a}, p. 61).\footnote{``a theory recently developed by Bhabha and Heitler and by Carlson and Oppenheimer, on the basis of quantum electrodynamics, leads to predict for high energy electrons a behaviour similar to the one that has been experimentally observed for soft rays. According to this theory, the main interaction between electrons and matter consists in the emission of a bremsstrahlung gamma radiation; the gamma quanta, in turn, undergo a process of materialisation giving rise to a couple of electrons, which emit gamma rays again, and so on. The showers are created by the alternate succession of those two processes of bremsstrahlung radiation and materialisation of energy; the soft local radiation is thus made of both positive and negative electrons and gamma rays''.} 

Then the question which still needed to be answered was: what about the nature of the penetrating rays? At the same Conference, Rossi thoroughly discussed the question. The hard corpuscles could not be electrons as they 
behaved differently from what Bhabha and Heitler theory predicted, 
though a large part of them had energies similar to those of 
the soft corpuscles. As Rossi said, ``Difficilmente quindi sembra 
di poter sfuggire alla conclusione che i corpuscoli molli e duri 
siano particelle di diversa natura'' (cf. \cite{Rossi1938a}, p. 62),\footnote{``It seems difficult to avoid the conclusion that the soft and hard corpuscles are particles of different nature''.} and he added 
that ``Poich\'{e} siamo stati condotti ad identificare i corpuscoli 
molli con elettroni, verrebbe fatto di pensare che i corpuscoli 
duri potessero essere protoni'' (cf. \cite{Rossi1938a}, p. 62).\footnote{``As we have been led to identify the soft corpuscles with electrons, we may think that the hard corpuscles could be protons''.} Such an hypothesis could not be excluded yet but, ``qualora successive esperienze conducessero ad escludere la presenza di protoni nella radiazione cosmica, 
non rimarrebbe, sembra, altra possibilit\`{a} se non quella di 
ammettere l'esistenza di corpuscoli finora sconosciuti'' (cf. \cite{Rossi1938a}, p. 63).\footnote{``if further experiments led to exclude the presence of protons in cosmic radiation, the only possibility left would be to accept the existence of corpuscles as of now unknown''.} According to Rossi, some observations carried out in 
a cloud chamber by Seth H. Neddermeyer and Carl Anderson seemed 
to suggest the presence of corpuscles which could not be identified 
with protons or electrons, and these two authors had proposed 
that the penetrating particles were particles of unitary charge 
with a mass larger that the mass of the electron and smaller 
than that of the proton \cite{NeddermayerAnderson1937}. Rossi concluded his talk with these words ``Ulteriori esperienze appaiono per\`{o} necessarie perch\'{e} si possa valutare l'attendibilit\`{a} di questa ipotesi'' (cf. \cite{Rossi1938a}, p. 64).\footnote{``Further experiments seem however to be necessary in order to determine the reliability of this hypothesis''.} As a matter of fact, Neddermeyer and Anderson had added 
at the end of their paper that an excellent experimental proof of 
the existence of such new particles had just been given, during 
a Meeting of the American Physical Society, at the end of April, 
by Jabez C. Street and Edward C. Stevenson \cite{StreetStevenson1937a}. The latter scientists sent their complete results for publication in October 
1937 and their paper was published in November 1937: doubts 
were no longer possible, there was a new particle, which was 
to be called ``mesotron'' or ``yukon'' (see the next paragraph to understand this apparently strange name), and later muon or $\mu$-meson \cite{StreetStevenson1937b}.

As Hideki Yukawa's relativistic theory of nuclear forces -- i.e. his 
theory of the interactions of heavy particles in nuclei -- predicted 
the existence of charged particles of mass intermediate between 
those of the electron and of the proton,  the mesotrons were immediately 
identified with Yukawa's particles \cite{Yukawa1935}. \footnote{The theory of Yukawa was ignored for various years. Rephrasing Yukawa's theory in present terms one can state a parallel between (quantized) Maxwell electromagnetic theory with interaction of infinite range  (Coulomb potential $\frac{1}{r}$) mediated by photon (mass equal to zero) and Yukawa's theory of nuclear (strong) interaction of finite range (Yukawa potential $\frac{e^{-kr}}{r}$, range $\frac{1}{k}$) mediated by a particle with mass $m$ different from zero ($\frac{1}{k} = \frac{\hbar}{mc}$). When experimental observations gave hints for the possible existence of a particle with a mass greater than the mass of the electron and smaller than the mass of the proton, the physicists' attention was drawn to Yukawa's theory and the mesotron was regarded as the particle which mediated nuclear interactions.}  They were thus also called ``yukons'' by a few scientists. Hans Euler and Werner Heisenberg then discussed in 1938 ``the hypothesis that the hard component of cosmic rays consisted of mesotrons 
produced in the upper layers of the atmosphere by primary electrons 
or photons and then disintegrating, as predicted by Yukawa's 
theory [\dots ] with a life-time of the order of $10^{-6}$ secs.'' (cf. \cite{Rossi1938b}, p. 993). Such an hypothesis led to an anomalous 
attenuation in the atmosphere due to the decay of mesotrons in 
flight \cite{EulerHeisenberg1938}. As a matter of fact, in Asmara, in 1933, Rossi and De Benedetti had been the first to observe such an anomalous effect, as they had 
detected a decrease in the intensity of cosmic rays passing through 
the atmosphere decisively larger than what was expected. They 
could not explain the phenomenon at that time, but Rossi immediately 
interpreted it in 1938, after the publication of Euler and Heisenberg's 
paper \cite{Rossi1938b}. 

It is worth pointing out that at that time, in 1938, Rossi was in Copenhagen, 
as he had just been forced to leave Italy because of 
the fascist racial laws. After short periods in Copenhagen and 
Manchester, Rossi went to the United States, working first at 
Cornell University, then at Los Alamos, and finally, from 1946 onwards, 
at the Massachusetts Institute of Technology (MIT). His departure 
was of course an invaluable loss for Italian physics.

\subsection{Cosmic-ray studies in Italy during World War II}

Besides Rossi, a few other physicists who had been with him in 
Arcetri, had also started working on cosmic rays. Gilberto Bernardini 
and Giuseppe Occhialini, in particular, both played an important 
role in the field, though their scientific careers were very 
different one from the other: Bernardini mainly spent the 1930s 
and the 1940s in Italy, and he significantly contributed to the 
development of cosmic ray researches within the country, while 
Occhialini spent many years abroad, where he gave 
fundamental contributions. We have already discussed Occhialini's 
working with Blackett on cloud chambers in Cambridge, and we 
will see more about his work in the following sections. 

As for Bernardini, he carried out his first researches 
on cosmic rays in the early 30s in Florence, where he collaborated 
with Sergio de Benedetti and Daria Bocciarelli. He studied the 
typical cosmic ray questions of those years, as for instance 
the absorption of the penetrating radiation at different zenithal 
inclinations, the influence of the earth magnetic field on the 
penetrating radiation, and cosmic ray showers (see in particular \cite{BernardiniDeBenedetti1933,BernardiniBocciarelli1934a,BernardiniBocciarelli1934b,BernardiniBocciarelli1935a,BernardiniBocciarelli1935b,BernardiniBocciarelli1935c}). He was assigned in 1937 a chair 
in Camerino, but a few months later, in 1938, he already moved 
to the University of Bologna. However, as the Institute of Physics 
in Bologna was not suitably organised for extensive research 
activity, Bernardini spent several days a week in Rome, where 
the scientific activity went on in spite of the war (cf. \cite{Amaldi1946}). Several 
young physicists - Bernardo Nestore Cacciapuoti, Oreste Piccioni, Ettore Pancini, Mariano Santangelo, Eolo Scrocco, Marcello Conversi -- joined Bernardini and worked with him on cosmic rays. 

At that time, as we have seen, the mesotron had just been discovered 
and his properties were thoroughly studied. In particular, several 
experiments indicated that mesotrons were unstable, with a life 
time of about $10^{-6}$ seconds. According to Yukawa's theory, in each disintegration process, an electron was supposed to be produced and the emission of a neutrino was also postulated, in order to fulfil the requirements of the conservation laws. The electron was expected to get, on the average, half 
of the total energy of the mesotron (see \cite{Rossi1940}, p. 469). The number of electrons accompanying the mesotron beam in the atmosphere was therefore expected to be increased because of the mesotron decay, as compared to the number of electrons in a condensed material. In other words, as Bernardini wrote in 1939, a consequence of the theory was that ``la componente elettronica o molle, presente nell'atmosfera, da almeno 2000 
m di altezza fino al mare, \`{e} da considerarsi come una radiazione 
secondaria di quella mesotronica e appunto prodotta da quest'ultima 
in conseguenza del processo di disintegrazione proprio ai mesotroni'' (see \cite{Bernardini_et_al1939}, p. 809).\footnote{``the electronic or soft component in the atmosphere, from an altitude of at least 2000 m to sea level, must be regarded as a secondary radiation of the mesotronic component, precisely produced by the latter as a consequence of the disintegration process of mesotrons''.} Bernardini 
also pointed out that ``the striking confirmations which the instability 
hypothesis has received [...] have generated a strong confidence 
that the `disintegration electrons' can be found in some way'' (see \cite{Bernardini_et_al1940}, p. 1018).  However, only indirect evidence had been brought forward up to that moment, and ``a direct experimental confirmation of their  existence is still wanting'' (see \cite{Bernardini_et_al1940}, p. 1018). Bernardini and his collaborators in Rome decided to follow ``a more direct line of attack'' by measuring the intensities of the soft and hard components 
``first at sea level in free air and then at a certain height 
above sea level under a layer of dense material having, as closely 
as possible, the same atomic number as air and equivalent in 
stopping power to the air layer between the two altitudes'' (see \cite{Bernardini_et_al1940}, p. 1019).\footnote{In those years, the physicists in Rome carried out various experiments and published several papers on the subject. See for intance \cite{Cacciapuoti1939}; \cite{BernardiniCacciapuotiFerretti1939} (where the authors describe the experiences they carried out in the Basilica 
di Massenzio in Rome); \cite{BernardiniCacciapuotiFerretti_et_al1939}.} They also compared ``the Rossi curves for 
small showers under the same conditions'' and thoroughly studied 
``the increase of the soft component with increasing altitude'' (see \cite{Bernardini_et_al1940}, p. 1018). Their results surprisingly indicated that the ``soft radiation observed at sea level is not entirely due to secondary processes of the mesotron'' (see \cite{Bernardini_et_al1940}, p. 1018). They wrote in 1939 that ``Dai risultati sembra lecito poter concludere, contrariamente a quanto \`{e} generalmente ammesso, che la radiazione elettronica che giunge al livello del mare \`{e}, in considerevole misura, costituita dal residuo della componente elettronica primaria'' (see \cite{BernardiniCacciapuotiFerretti_et_al1939}, p. 1010).\footnote{``From the results, it seems possible to conclude, in contrast with what is generally thought, that the electronic radiation arriving at sea level is, in a large part, composed of the residue of the primary electronic component''.}  Bernardini and his group also carried out experiments on this subject in underground cavities and, in 1941, they suggested two possibilities ``o che gli elettroni osservati a grande profondit\`{a} non siano tutti secondari dei mesotroni, ma che una parte di essi siano generati da una ulteriore radiazione non ionizzante; oppure che l'interazione, 
con conseguente produzione di secondari, tra mesotroni e materia, 
sia pi\`{u} complessa di quanto ammette la teoria attuale'' (cf. \cite{Bernardini_et_al1941}, p. 321).\footnote{``either the electrons observed at great depths are not all secondaries produced by the mesotrons, but a part of them is generated by a further non ionising radiation; either the interaction between mesotrons and matter, with the resulting production of secondaries, is more complex with respect to what the current theory supposes''.}\footnote{Several underground experiments, intended to investigate various properties of the cosmic radiation, were carried out in those years and later in Italy and abroad (see for instance \cite{Amaldi1953}).} As a matter of fact, the relationship between the electronic and mesotronic components of cosmic rays could not be explained at that time and it was to remain an open question for a few more years, until the discovery of the $\pi$-mesons and of their decay modes. 

During the war time, Bernardini and a few other physicists, in 
Rome, also studied the positive excess in mesotron spectrum. 
Such an excess had already been examined in the early 1930s by 
Rossi\cite{Rossi_Nature1931} and by Lewis M. Mott-Smith \cite{Mott-Smith1932}, who had both tried to observe the deflection of cosmic particles by the magnetic field in an iron core. Rossi had used two coincident counters 
and magnetised iron bars interposed. Mott-Smith had obtained 
a negative result, and Rossi had found a very small effect, which 
pointed to an excess of the positive over the negative particles, 
but Rossi himself did not regard his result as a definite evidence. 
Cloud chambers experiments had then shown an approximately equal 
number of positive and negative particles \cite{Blackett1937}, but Louis Leprince-Ringuet in 1937 \cite{Leprince-Ringuet1937}, Haydn Jones in 
1939 \cite{Jones1939}, and Donald J. Hughes in 1940 \cite{Hughes1940} had found a positive excess of about 20 percent. Bernardini then proposed to ``repeat the experiment of Rossi with a somewhat improved triple coincidence 
arrangement'' (see\cite{BernardiniWick_et_al1941a}, p. 536; see also \cite{BernardiniConversi1940, BernardiniWick_et-al1941b}; figs. 10a-10b-10c). 
\begin{subfigures}
\begin{figure}
\centering
  \includegraphics[scale=0.5]{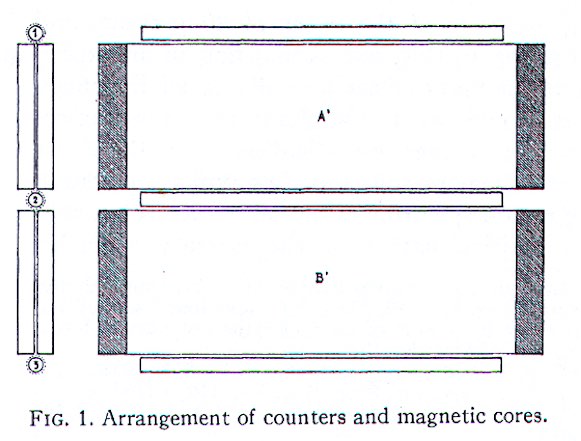}
  \caption{\label{fig:10A}} 
\end{figure}
\begin{figure}
\centering
 \includegraphics[scale=0.35]{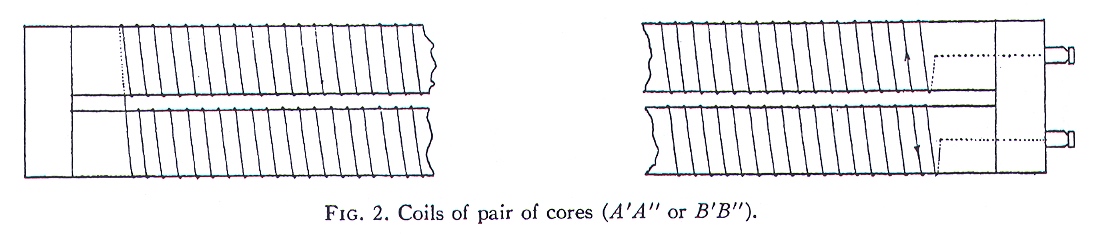}
  \caption{\label{fig:10B}} 
\end{figure}
\begin{figure}
\centering
 \includegraphics[scale=0.3]{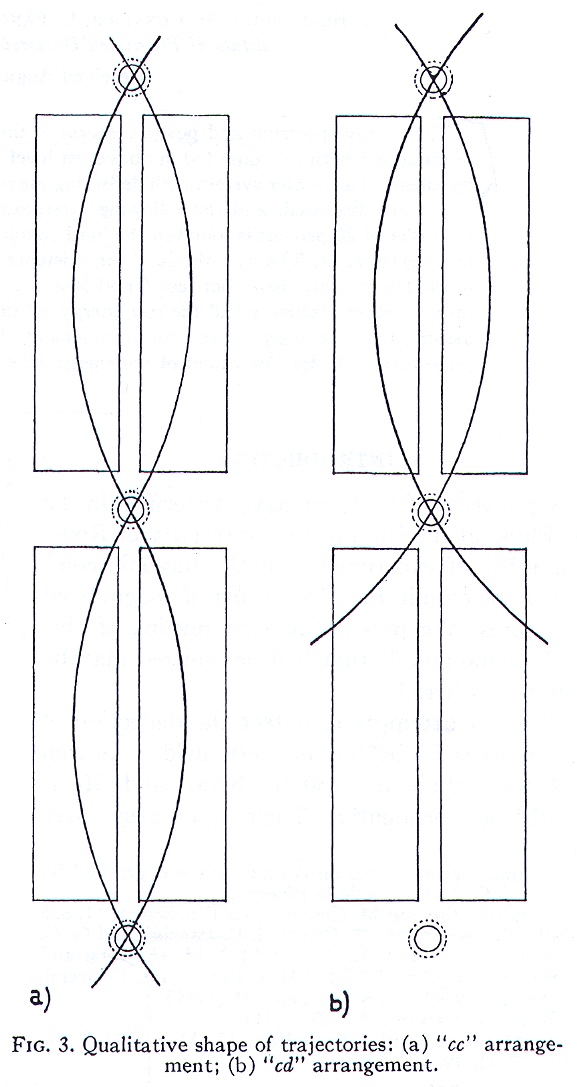}
  \caption{\label{fig:10C}The three figures 10a-10b-10c (reproduced from \cite{Bernardini_et_al1945} but similar to those given in \cite{BernardiniWick_et-al1941b}) schematically show the apparatus used by Bernardini and collaborators. It consists of three coincident counters and two pairs of iron cores $A'$, $A''$ and $B'$, $B''$. The field in the cores is parallel to the axis of the counters and has opposite directions in $A'$, $A''$ (similarly in $B'$, $B''$). The magnetic field is closed by iron bars applied at both ends of $A'$, $A''$ and $B'$, $B''$. Thus each pair of iron cores acts like a cylindrical magnetic convergent (``$c$'') or divergent (``$d$'') lens for positive ($c+$ or $d+$) or negative ($c-$ or $d-$) particles (cf. \cite{Bernardini_et_al1945}, p. 111).} 
\end{figure}
\end{subfigures}
Working in Rome and at Pian Ros\`{a} (Cervinia), at 3460 meters above sea level, the Rome 
group found out in 1941 ``a conclusive evidence in favour of the 
existence of a positive excess'' and immediately planned further 
experiments, as the method ``owing to its simplicity, seems well-suited 
for an investigation [\dots ] under conditions when the elaborate 
Wilson chamber technique is impossible'' (see \cite{BernardiniWick_et_al1941a}, p. 536). They performed 
more experiments in the years 1941-1943 and published several 
papers (see in particular \cite{BernardiniConversi1940, BernardiniWick_et_al1941a, BernardiniWick_et-al1941b, ConversiScrocco1943}). In an outline 
of their research work, published in the \textit{Physical Review} in 1945 \cite{Bernardini_et_al1945}, 
they explained that they had investigated the energy spectrum 
and positive excess of the hard component of cosmic rays for 
``cogent reasons'', as ``the positive excess in the meson spectrum 
is probably connected with the positive nature of the primary 
radiation'' and ``a study of the variation of the positive excess 
with height would probably be interesting and might throw some 
light on the process of creation of the mesons'' (see \cite{Bernardini_et_al1945}, p. 110). Both in Rome and at Pian Ros\`{a}, 
the Rome physicists had used the same experimental apparatus, 
i.e. ``a counter system with deflecting magnetised cores'' (see \cite{Bernardini_et_al1945}, p. 109, and fig. 10c). The latter, which were similar to the ones used by Rossi \cite{Rossi_Nature1931}, were to be called ``magnetic lenses'' in further experiments. Bernardini and his collaborators concluded in particular that ``a positive excess of the order of 20 percent is found in 
the hard component, in agreement with the results of other investigators'' 
and that ``the hypothesis of the existence of several types of 
mesons is not confirmed in the lower atmosphere'' (see \cite{Bernardini_et_al1945}, p. 109). In fact, the researches on the positive excess went on for several more years, as they could provide details about the composition of cosmic rays at different heights, and about the production 
of particles and other interactions which took place within cosmic 
rays at different altitudes \cite{Janossy1943, BridgeRossi1947, BridgeRossiWilliams1947}. In Rome, these experiments went on after the war: the researches were performed at higher and higher heights, but the same kind of apparatus -- the magnetic lenses -- were used for several 
more years.\footnote{See for instance the papers \cite{Quercia_et_al1948a, Quercia_et_al1948b, Ballario_et_al1948}. All the authors thank Bernardini for his assistance, his advice or for the stimulating discussions.}

Finally, during WWII, the Rome physicists also worked on the 
mesotron decay, by examining at first the anomalous attenuation 
of mesotrons in the atmosphere. Once again, they worked in Rome 
and around Cervinia (in Ch\^atillon and at Pian Ros\`{a}, at respectively 
500 and 3460 meters above sea level), and they compared the vertical 
intensity of mesotrons at these two altitudes \cite{Ageno_et_al1939, Ageno_et_al1940}. They concluded that ``The anomalously large absorption 
of mesotrons in air has been confirmed'' and that ``If the results 
obtained are interpreted according to the hypothesis of the instability 
of the mesotron, they are found to be consistent with the assumption 
of a proper lifetime of 4 or 5 microseconds for the mesotron'' (see \cite{Ageno_et_al1940}, p. 945). As there were discrepancies between the results obtained at that time by several authors, Bernardini and his group carried 
out further experiments \cite{BernardiniCacciapuoti_et_al1941} and they found values for the lifetime 
of the mesotron which were in agreement with some authors (Rossi 
and David B. Hall \cite{RossiHall1941}; H.Victor Neher and H. Guyford Stever \cite{NeherStever1940}) and in disagreement with others (Walter M. Nielsen et al. \cite{Nielsen_et_al.1941} in particular). As they wrote in 1942, their own results and the comparison with other authors showed how ``la precisione 
reale raggiunta fino a oggi [\dots ] in base all'assorbimento anomalo, 
sia molto mediocre'' (see \cite{BernardiniCacciapuoti_et_al1942}, p. 98).\footnote{``the real precision achieved as of now [\dots] on the basis of the anomalous absorption, is very mediocre''.} As a matter of fact, it may be worth underlining that the measures of the meson mean life based on the anomalous absorption were only indirect proofs of the disintegration of mesons. Direct proofs of the instability of mesotrons had been supplied by cloud chambers photographs obtained by Evan James Williams and G.E. Roberts in 1940 \cite{WilliamsRoberts1940}, and by  Ralph P. Shutt, De Benedetti, and Johnson in 1942 \cite{ShuttDebenedettiJohnson1942}. Moreover, by 1939, a direct measure of the mesotron mean life had also 
been tried by Carol G. Montgomery and his collaborators, who studied the delayed coincidences between two layers of Geiger counters separated by a layer of lead intended to stop the mesotrons \cite{Montgomery_et_al1939}. The idea was to measure the time which elapsed from the stopping of the mesotron in matter up to its disintegration in an electron or a positron. This first 
direct measure did not give a good result because of too many 
spurious events. In 1941, two new direct experiments, one by 
Franco Rasetti \cite{Rasetti1941a, Rasetti1941b}, who was in Quebec at that time, and the other one by Pierre Auger, Roland Maze and Robert Chaminade in France, were proposed \cite{Auger_et_al1941, MazeChaminade1942}. Then, in the autumn 1941, two of Bernardini's 
collaborators, Marcello Conversi and Oreste Piccioni, without knowing 
about the French experiment, decided to try a ``direct'' experiment 
too. Their work, finished in February 1943, was published in \textit{Nuovo Cimento} \cite{ConversiPiccioni1944a, ConversiPiccioni1944b}, but it was not published in the\textit{ Physical Review} until 1946 because of the wartime\cite{ConversiPiccioni1946a, ConversiPiccioni1946b}. Their experiment, 
which was ``performed by counting delayed coincidences between 
the impinging low energy mesons and the decay electrons'' (cf. \cite{ConversiPiccioni1946a}, p. 859) marked a significant improvement in the value of the meson mean life and their results ($2.3 \pm 0.15$ $\mu$s) were in excellent agreement with similar ones obtained at about the same time at Cornell University by Rossi and Norris Nereson ($2.15 \pm 0.07$ $\mu$s) \cite{RossiNereson1942, RossiNereson1943}. The apparatus designed by Conversi and Piccioni was made of three layers of Geiger counters (fig. 11). 

\begin{figure}

\centering

\includegraphics[width=\textwidth]{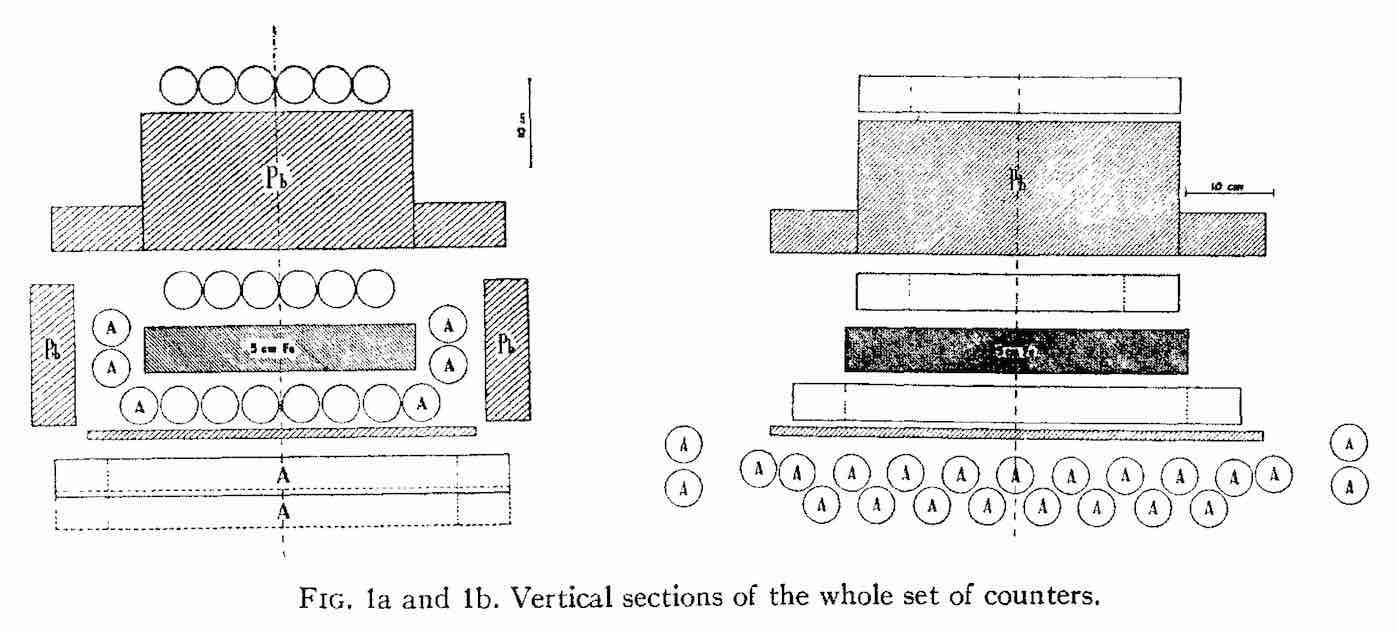}

\caption{The experimental arrangement of the apparatus designed by Conversi and Piccioni. It shows, in a vertical section, the disposition of the four groups of counters, the lead screens, and the iron absorber (reproduced from \cite{ConversiPiccioni1946b}, p. 875).}\label{fig:11}

\end{figure}

The first two layers, separated from one another by a layer of lead, detected the arrival of a mesotron of suitable energy. An absorber 
formed by an iron plate was inserted between the second and the 
third layer of counters. The mesotrons stopped in the iron absorber 
decayed, giving an electron, and a delayed coincidence was thus 
measured by the third layer of counters. A fourth group of counters formed a so called ``anti-coincidence'' layer, which was intended to eliminate the spurious coincidences. ``Of course -- as Piccioni later wrote -- when I came to know the technical electronic development done in US during the war I was overwhelmed. But it was rewarding to see that the concepts of short rise times, delays with monostable circuits and binary scalers circuits similar to (though better than) ours were also important features of the post war electronic inventory'' (cf. \cite{Piccioni1987}, p. 178). And in fact, although the electronic technique was not highly sophisticated with respect to US, Conversi and Piccioni's experiment, achieved in the very difficult Italian war conditions, was regarded as remarkable.

Besides Rome, the only other place in Italy where significant 
cosmic-ray researches went on during WWII was Milan.\footnote{A few 
isolated papers were published in some other universities, such as Messina (by Salvatore Patan\`e and Beltramino Panebianco\cite{Patane1941, PatanePanebianco1941a, PatanePanebianco1941b, Patane1942}), Pavia (by Rita Brunetti and Zaira Ollano \cite{BrunettiOllano1941, BrunettiOllano1942}) and Torino (Giuseppe Lovera \cite{Lovera1943a, Lovera1943b}), but no systematic, regular work was carried out in these universities.}  Giuseppe Cocconi had graduated there in 1937 and he had then been several times 
in Rome, in 1938-1939, working on the mesotron decay with Fermi, 
Rasetti and Bernardini (cf. \cite{Amaldi1979}, p. 188). Back to Milan, Cocconi 
went on working on cosmic rays, alone or with Vanna Tongiorgi -- who became  his wife -- and Andrea Loverdo. Cocconi and his collaborators studied the secondaries of the penetrating radiation and the equilibrium of the soft and hard components at sea level \cite{CocconiTongiorgi1939a, CocconiTongiorgi1939b, CocconiTongiorgi1940a}, the neutrons within cosmic rays \cite{Cocconi1939}, the mesotron mean life and the extensive air showers \cite{Cocconi1940, CocconiTongiorgi1940b}. For their experiments, 
the Milan group mostly used sets of Geiger-M\"uller counters, but 
a cloud chamber, which was designed in 1939 and finished in 1941, 
was also available \cite{Polvani1941}.\footnote{Giovanni Polvani, Director of the Milan Institute of Physics got the funds for the construction of the 
cloud chamber from C.N.R. and from the Societ\`{a} Edison.} 
In fact, Cocconi used it only for a few experiments \cite{Cocconi1941a, Cocconi1941b, Cocconi1942}, as he left Milan and spent a long time, in 
1942 and 1943, at Passo Sella (Dolomiti), carrying out with Vanna 
Tongiorgi, and for a limited time by Loverdo as well, at 2200 meters above sea level, an important set of experiments which were to be published in the \textit{Physical Review} at the end of the war \cite{CocconiLoverdoTongiorgi1946a, CocconiLoverdoTongiorgi1946b, CocconiTongiorgi1946a, CocconiLoverdoTongiorgi1946c, CocconiTongiorgi1946b} (see also \cite{CocconiLoverdoTongiorgi1943a, CocconiLoverdoTongiorgi1943b, CocconiLoverdoTongiorgi1944a, CocconiLoverdoTongiorgi1944b}). It is worth mentioning that Cocconi also discussed in 1941 the protonic nature of the primary radiation, which had been suggested by Thomas H. Johnson and J. Griffiths Barry in 1939 \cite{JohnsonBarry1939} on the basis of the altitude dependence of the east-west effect. Protons were then regarded, in 1941, as the only component of the primary cosmic radiation by Marcel Schein, William P. Jesse, Ernesto O. Wollan, who showed that there is a hard component increasing to the highest altitudes, and by William Francis Gray Swann on the basis of other general considerations\cite{Schein_et_al1941, Swann1941a, Swann1941b}. After the war, in 1946, Cocconi was assigned a chair at the University of Catania, and he brought there his cosmic-ray experiments. Then he left for New-York, where he started working at Cornell University in 1947. 

In the meanwhile, at the end of the war, a new group of young physicists started carrying out researches on cosmic rays in Milan: the group was composed by Giorgio Salvini, Antonino Mura and Guido Tagliaferri, who were later joined by Antonio Lovati.

\section{Cosmic rays after World War II: the rebirth of Italian physics}

``A disaster''. With this single word, Italian scientists described 
the condition of Italian physics at the end of WWII. Many Italian 
physicists  -- like Fermi, Rossi, Pontecorvo and Occhialini -- had left the country for political reasons. Some of the physics institutes -- Pisa and Palermo for instance -- had been severely bombed, and in several places the scientific apparatus had been stolen or damaged. Finally, many physicists 
had been involved in the war, and scientific activity had almost 
completely stopped, even in the places, like Padua, which had 
not been damaged. At the end of the war, scientists 
started coming back from the war experience, but they knew nothing 
about the recent results and techniques. Moreover, the funds 
for scientific activity were very poor. Nicol\`{o} Dallaporta, 
who had been invited to work in Padua by Antonio Rostagni -- 
the new professor of experimental physics after Rossi's departure 
-- describes the life in the first post-war years in Padua with 
these words: ``we were all of us widely ignorant concerning the developments of physics during the six years period of the war, as few journals had been available in that time. Thus, the only thing to begin with was reading as much as possible in order to recover time lost and get able to choose a field of research suitable for the rather poor conditions we had to face, at least for some time; with an eye open, however, on future possibilities of growth, as it appeared reasonable to hope in an adequate increase of both the staff members and the financial support. There was no heating in the building during this first winter'' (cf. \cite{Dallaporta1989}, p. 534).

Only a very few places -- Rome and Milan, as we have seen -- had 
kept some researches going on but, as Edoardo Amaldi points out 
about the situation in Rome, if on the one hand ``era apparso 
evidente che nel campo della ricerca fondamentale ci eravamo 
mantenuti al corrente in modo soddisfacente durante tutta la 
guerra'',\footnote{``it was clear that, as for fundamental research, we had kept ourselves satisfyingly informed during the whole wartime''.} on the other hand ``eravamo invece rimasti estremamente 
indietro per quanto riguardava le tecniche sperimentali'' (cf. \cite{Amaldi1979}, p. 206).\footnote{``we had stayed far behind as for the experimental techniques''.} We will examine at first what went on in Rome right after the war, as it was quite a peculiar situation with respect of the rest of Italy. We will then turn to Occhialini's further contributions to cosmic ray researches, and we will finally discuss the general rebirth of Italian physics and the further developments of cosmic ray physics in Italy.

\subsection{Cosmic rays in Rome in the first post-war years: Pancini-Piccioni-Conversi experiment}

We have seen that, in Rome, the physicists had worked on the one 
hand on the positive excess by using magnetised iron cores, and that they had carried out on the other hand experiments on the mean life of the mesotron by using delayed coincidences systems. Conversi and Piccioni, who had performed these latter experiments, were later joined by Ettore Pancini, and the three physicists decided to use their delayed coincidences apparatus together with the magnetic lenses system, in order to carry out a separate analysis of the behaviour of positive and negative mesons in matter (fig. 12). In particular, in 1940, Sin-itiro Tomonaga and Gentaro Araki \cite{TomonagaAraki1940} had pointed out that ``because of the Coulomb field of the nucleus, the capture probability for negative mesons at rest would be much greater than their decay probability, while for positive mesons, the opposite should 
be the case'' (cf. \cite{ConversiPanciniPiccioni1947}, p. 210). It was thus expected that ``if this is true, then practically all the decay processes which one observes should be owing to positive mesons'' (cf. \cite{ConversiPanciniPiccioni1947}, p. 210). Conversi, Pancini and Piccioni carried out a first set of experiments in 1945, registering the decay electrons in 
an iron absorber, and they obtained results showing ``the greatly 
different behaviour of negative and positive mesons, so that 
the prediction of Tomonaga and Araki seems to be confirmed experimentally'' (cf. \cite{ConversiPanciniPiccioni1945}, p. 232). The three young Italian physicists then decided to use as absorber a lower atomic number material, and they chose carbon. It seems that the main reason of this new experiment was to observe high energy gamma rays emitted after the nuclear capture of negative 
mesons \cite{Salvini2004}. The result they obtained turned out ``to be quite inconsistent with Tomonaga and Araki's prediction'' (cf. \cite{ConversiPanciniPiccioni1947}, p. 210), as they surprinsingly found out that 
both positive and negative mesons mainly decayed when the carbon 
layer was inserted: the expected nuclear interactions did not seem to take place. The astonishing result was sent in December 1946 to the \textit{Physical Review} and published in February 1947. 

\begin{figure}

\centering

\includegraphics[width=\textwidth]{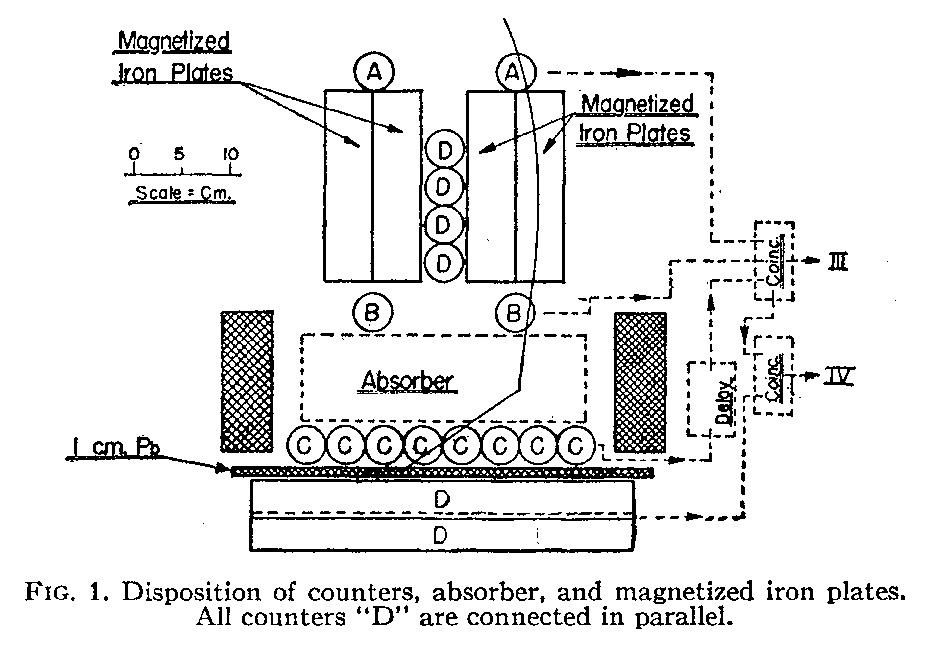}

\caption{The figure of the apparatus used by Conversi, Pancini, and Piccioni is reproduced from \cite{ConversiPanciniPiccioni1947}, p. 209.}\label{fig:12}

\end{figure}

Amaldi directly wrote to Fermi, to inform him about these results, 
and Fermi, who discussed the matter with Edward Teller and Victor Weisskopf, concluded that ``If the experimental results are correct, they would necessitate a very drastic change in the forms of mesotron interactions'' (cf. \cite{Fermi_et_al1947}, p.315; see also \cite{FermiTeller1947}). As we will see, the first experimental answer to 
this open question was to be provided by Cecil F. Powell and 
his group a few months later. It is worth pointing out that Luis 
W. Alvarez, on accepting the Nobel Prize in 1968, wrote that 
``As a personal opinion, I would suggest that modern particle 
physics started in the last days of World War II, when a group 
of young Italians, Conversi, Pancini, and Piccioni, who were hiding 
from the German occupying forces, initiated a remarkable experiment. 
In 1946, they showed that the `mesotron' which had been discovered 
in 1937 [...] was not the particle predicted by Yukawa as the 
mediator of nuclear forces, but was instead almost completely 
unreactive in a nuclear sense. Most nuclear physicists had spent 
the war years in military-related activities, secure in the belief 
that the Yukawa meson was available for study as soon as hostility 
ceased. But they were wrong'' (cf. \cite{Alvarez1972}, p. 241-2).

\subsection{Occhialini in Bristol and Bruxelles}

After his collaborating with Blackett in Cambridge, Giuseppe Occhialini had moved to S\~{a}o Paulo in Brazil, then, at the 
end of 1944, he went to the Wills Laboratory in Bristol UK in order to 
collaborate with Cecil F. Powell. 

Powell had been working from 
1939 on nuclear emulsions, i.e. photographic emulsions of very 
high silver concentration thickly coated on glass backing. When 
an ionising particle passed through the emulsion, it left behind 
a number of silver bromide crystals so altered that upon development 
they appeared as rows of black grains of colloidal silver and 
thus identified the track of the particle. The more strongly 
ionising the particles, the more numerous are these grains and 
thus, as a swift particle has less power of ionising than a slow 
one, the greater the speed of a particle, the greater the distance 
between the grains. Moreover, the greater the initial energy 
of the particle, the longer the resulting track. As there are 
relationships accurately connecting all these quantities, it 
is usually possible to identify the particle and its energy. This 
way of recording particles tracks, discovered by Antoine Henri Becquerel 
in 1896 \cite{Becquerel1896}, had been in use from the early 20$^{\mbox{th}}$ century to study radioactive radiation, but it had not been sensitive and reliable enough until the improvements that Powell and his collaborators introduced.\footnote{For a description of the photographic method, see \cite{Breiser1952, Goldschmidt-Clermont1953}.} They improved the 
treatment of the material, the research technique and the optical 
equipment for analysing the tracks. In the meantime, new emulsions, 
more concentrated and sensitive, were produced by Ilford, so 
that in those years the photographic method was to become one 
of the most precious tools of particle physics. 

In particular, a few months after Pancini, Piccioni, and Conversi 
experiment, in May 1947, Powell, Occhialini and their collaborators 
in Bristol presented two events, found on photographic plates 
exposed at the Pic du Midi (2800 meters altitude), both showing 
a meson coming to the end of its range in the emulsion and producing 
a secondary particle, which was a second meson \cite{Lattes_et-al1947}. As they pointed out at the beginning of their paper, ``It is convenient to apply 
the term `meson' to any particle with a mass intermediate between 
that of a proton and an electron'' (cf. \cite{Lattes_et-al1947}, p. 694). Very cautiously, they wrote that ``It is therefore possible that our photographs 
indicate the existence of mesons of different mass'' (cf. \cite{Lattes_et-al1947}, p. 696) and they concluded, referring to Pancini, Piccioni and 
Conversi experiment and to Fermi, Teller and Weisskopf's suggestions 
that ``Since our observations indicate a new mode of decay of 
mesons, it is possible that they may contribute to the solution 
of these difficulties'' (cf. \cite{Lattes_et-al1947}, p. 697). 

In their following paper, published in October 1947, Powell and 
his collaborators extended their observations to plates exposed 
at 5500 meters in the Bolivian Andes \cite{LattesOcchialiniPowell1947}. They found ``forty examples of the process leading 
to the production of secondary mesons'' and in eleven of these 
cases, the secondary particle was ``brought to rest in the emulsion 
so that its range can be determined'' (cf. \cite{LattesOcchialiniPowell1947}, p. 453). There were now no doubts. The measurements made on the tracks ``provide evidence for the existence of mesons of different mass'' (cf. \cite{LattesOcchialiniPowell1947}, p. 453). As a matter of fact, in the meanwhile, in June 1947, at the Shelter Island Conference, Hans Albrecht Bethe, and Robert Eugene Marshak had already proposed a two-meson explanation of Conversi, Pancini and Piccioni effect \cite{MarshakBethe1947}.\footnote{Let's point out that Shoichi Sakata and Takesi Inoue had already proposed two types of 
mesons in 1946 \cite{SakataInoue1946}, but it seems that 
only a few scientists had seen it and that Marshak and Bethe 
did not know about this paper (see D. H. Perkins \cite{Perkins1989}).} Powell's new observations clearly confirmed such an hypothesis. In particular, there was ``good evidence for the production of a single homogeneous group of 
secondary mesons, constant in mass and kinetic energy. This strongly 
suggests a fundamental process'' and the Bristol physicists added 
that it was ``convenient to refer to this process in what follows 
as the $\mu$-decay'' (cf. \cite{LattesOcchialiniPowell1947}, p. 454-5). They proposed to ``represent the primary mesons by the symbol $\pi$, and the secondary by $\mu$'' (cf. \cite{LattesOcchialiniPowell1947}, p. 455). Their observations suggested that ``the heavier $\pi$-mesons suffer spontaneous decay with the emission of the lighter $\mu$-mesons'' (cf. \cite{LattesOcchialiniPowell1947}, p. 492). Powell and his collaborators also discussed the origin of the slow mesons which had recently been observed as producing nuclear disintegrations.  Such mesons had been observed in the previous months by Donald H. Perkins and by Powell and Occhialini as well \cite{Perkins1947, OcchialiniPowell1947}. The latter provisionally referred to these mesons as ``$\sigma$-mesons'' but it was quite clear that ``$\pi$-, and a large proportion of the $\sigma$-mesons are, respectively, positively and negatively charged particles of the same type''(cf. \cite{LattesOcchialiniPowell1947}, p. 492). The Bristol physicists showed that these particles not only produced nuclear disintegrations, but that they could be ``generated in the disintegration of nuclei by cosmic ray particles of great energy'' (cf. \cite{LattesOcchialiniPowell1947}, p. 489). In addition, by 
taking into account their own results as well as ``those obtained 
in delayed coincidences and Wilson chamber experiments'', Powell 
and his collaborators wrote that all the experiments performed 
up to that moment could ``be explained on the assumption that 
the greater part of the mesons observed at sea-level are $\mu$-mesons 
formed by the decay in flight of $\pi$-mesons''  and ``that positive 
and negative $\pi$-mesons are short-lived, with a mean life-time 
in the interval from 10$^{-6}$ to 10$^{-11}$ sec.'' (cf. \cite{LattesOcchialiniPowell1947}, p. 492).

At the 1982 \textit{Paris Colloque on the History of Particle Physics}, 
Charles Peyrou thus summarised Lattes, Occhialini and Powell's conclusions: 
``There are two mesons: $\pi$ and $\mu$. The $\pi$ decays into a $\mu$ and 
a unique neutral particle because the $\mu$ emitted in the decay 
has always the same range. The best interpretation of the Conversi, 
Pancini and Piccioni result is to assume the $\pi$'s have a strong 
interaction and are produced in nuclear interactions (in particular 
the ones made by the primary cosmic rays at  the top of the atmosphere). 
They decay very quickly into $\mu$'s which have very weak interactions 
with nuclei and constitute the bulk of the cosmic ray components 
at sea-level i.e. they are mesotrons. Mesons stopping in emulsion 
and giving rise to a star (nuclear reaction) are negative $\pi$'s 
obeying the Tomonaga and Araki prescriptions. Mesons stopping 
in emulsions and doing nothing are $\mu^{\pm}$ (decays 
electrons could not be seen in the emulsion of those days)'' (cf. \cite{Peyrou1982}, p. 21).

In 1950 Powell was to be awarded with the Nobel Prize for ``\textit{his development of the photographic method for the studying of nuclear 
processes and his discoveries concerning the mesons}''. For 
the second time, after his working with Blackett, Giuseppe Occhialini 
had thus thoroughly collaborated to a Nobel Prize discovery. Bruno 
Pontecorvo summarised the singular situation of Occhialini with this sentence, written on the occasion of the Symposium organised to celebrate the twentieth anniversary of Occhialini's coming back in Italy:\footnote{The proceedings of the Symposium, held in Milan on 10 October 1968, are published in \cite{SimposioOcchialini1969}.}  ``It would be very easy for me not to offer a toast to Peppino, but to any physicist, more or less in this way: I raise my glass with the hope that you can collaborate with Occhialini in some experiments. It is practically a certain way for you to win the Nobel Prize'' (cf. \cite{Vegni2006}, p. 124).

In the last months he spent in Bristol, Occhialini started collaborating with Constance Dilworth, who was to become one of his most brilliant collaborators. Constance also became Occhialini's wife in 1949. Together, they improved the technique of developing emulsions and studied new emulsions of high sensitivity. In particular, with the collaboration of Ron Payne, Occhialini and Dilworth devised in 1948 a new method to process thick emulsions. This was crucial because ``in the application of the photographic method to research in cosmic-ray and nuclear physics, the need has always 
been felt for emulsions of thickness comparable with the range 
of the particles to be recorded'' (cf. \cite{OcchialiniDilworthPayne1948}, p. 102). However, the use of thick emulsions had always been ``greatly restricted [\dots ] by the difficulties met in processing them'' (cf. \cite{OcchialiniDilworthPayne1948}, p. 102). As developers only 
penetrate by a rather slow diffusion process from the surface, 
the central problem was to achieve an even development throughout 
the thickness and over the whole area of the plate. A high contrast 
and absence of distortion were also highly required. As for distortion, 
it was essential for instance to avoid strong gradients of concentration 
or temperature in the emulsion or close to it, and to reduce 
to a minimum the osmotic pressure effects. A simple method to 
achieve an even development of thick emulsions was to choose 
a slow developer, and such a method was regarded in the early 
1950s as successful for thicknesses up to 400 microns. As for 
Dilworth, Occhailini and Payne, their method was based on the 
fact that ``the rate of permeation has a lower temperature coefficient 
than the rate of development'' (cf. \cite{OcchialiniDilworthPayne1948}, p. 102). In practice, the plates were first soaked in water and then in a cold developer: 
no appreciable development occurred but the developer was allowed 
to diffuse evenly throughout the thickness of the emulsion. The 
plates were then ``protected from a further increase in the concentration 
of the developing agent'' (cf. \cite{OcchialiniDilworthPayne1948}, p. 102), and the temperature was raised, so that the developer was to produce identical effects in all layers. After a suitable time, the temperature was lowered again 
to stop the developing effects. In 1948, Dilworth, Occhialini 
and Payne succeeded in processing 700 microns plates without 
difficulties. Their method, called ``temperature development'' 
(TD), was further improved and it became the most widely used 
technique by the early 1950s, as it proved successful for thicknesses 
up to 1200 and even 2000 microns.\footnote{For a further bibliography 
see \cite{Goldschmidt-Clermont1953}.} 

At the end of 1948, Occhialini and Dilworth left Bristol and 
went to the Universit\'{e} Libre of Bruxelles, as Occhialini had 
been invited there by Max Cosyns -- who was at the head of the 
Centre de Physique Nucl\'{e}aire -- to direct the photographic plates 
research laboratory. Then, one year later, Occhialini departed 
again, as he had been assigned a chair at the University of Genoa. 
In 1952 he finally moved to the University of Milan, where he 
was to hold the Chair of Fisica Superiore until the end of his 
academic career in 1983. Constance Dilworth joined him in Milan, 
as well as Livio Scarsi and Alberto Bonetti, who had both started 
collaborating with Occhialini in Genoa. 

\subsection{Cosmic rays as low-cost means to study particles physics 
in Italy after the war} 

As we have said, the general situation in Italy just after the war was very bad but, little by little, as the scientists started coming back to their research Institutes, scientific activity 
started again. At the national level, the Italian government 
was ``desideroso'', according to Amaldi, ``di riportare il CNR alla 
normalit\`{a}'' (\cite{Amaldi1979}, p. 194),\footnote{``willing to bring CNR to normality again''.} so that at the beginning of 
1945, the Chair of CNR was assigned to Gustavo Colonnetti, professor 
of  Scienze delle Costruzioni at the Politecnico di Torino. The 
Committee for Physics and Astronomy was formed at that time as 
well, and it was chaired by the physicist Eligio Perucca, who 
was also working at the Politecnico di Torino. 

As a first concrete result of this ``institutional'' renaissance 
of Italian phyiscs, a \textit{Centro di studio per la fisica 
nucleare} -- later called \textit{Centro di studio per la fisica 
nucleare e delle particelle elementari} -- was founded in 1945 in 
Rome, at the Istituto di Fisica Guglielmo Marconi. The Rome Institute 
of Physics thus obtained from CNR more substantial funds on a regular 
basis. This research center was expected to continue the researches 
carried out up to that moment, i.e. cosmic-ray and nuclear processes 
researches. As for this latter field, the experiments were performed 
at the Istituto Superiore di Sanit\`{a}, where a 1.1 milion volts 
accelerator had been built in 1937-38.\footnote{See the papers by Amaldi in \textit{Ricerca Scientifica}, who described every year, from 1945 to 1951, the scientific activity of the Centro di Studio \cite{Amaldi1945_1951}.} After the war, the construction of a 20 MeV betatron was also proposed but the project was abandoned by 1947, ``non solo perch\'{e} i mezzi a disposizione erano insufficienti, ma anche perch\'{e}'', according to Amaldi, ``non potevamo ancora contare sull'apporto dell'industria italiana, totalmente impegnata 
nei lavori inerenti la ricostruzione generale del Paese. [\dots] 
Tutto lo sforzo fu quindi concentrato sullo studio della radiazione 
cosmica'' (\cite{Amaldi1979}, p. 207).\footnote{``not only because the available means were not sufficient, but also because we could not count yet on the contribution of the Italian industry, which was totally committed to the general reconstruction of the Country. [\dots] The whole effort was thus concentrated on the study of the cosmic radiation''.}

Within this context, the Roman group of physicists designed and built at an altitude of 3500 metres, close to Cervinia, the so called ``laboratorio della Testa Grigia'', intended for the study of elementary particles within cosmic rays. The laboratory was inaugurated on the 11th January 1948 \cite{BernardiniLongoPancini1948, BernardiniPancini1950, Fidecaro1955} (figs. 13a and b). 

\begin{subfigures}
\begin{figure}
\centering
  \includegraphics[scale=0.6]{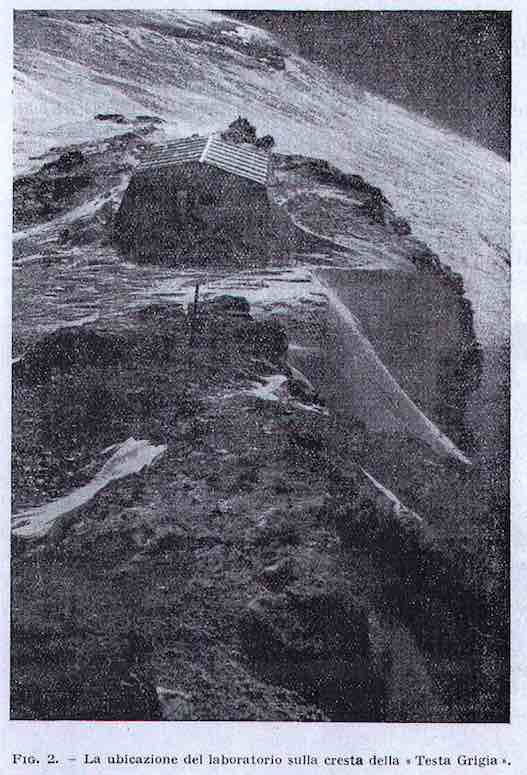}
  \caption{\label{fig:12bisA}A photograph of the Testa Grigia Laboratory, reproduced from \cite{BernardiniLongoPancini1948}.} 
\end{figure}
\begin{figure}
\centering
 \includegraphics[width=\textwidth]{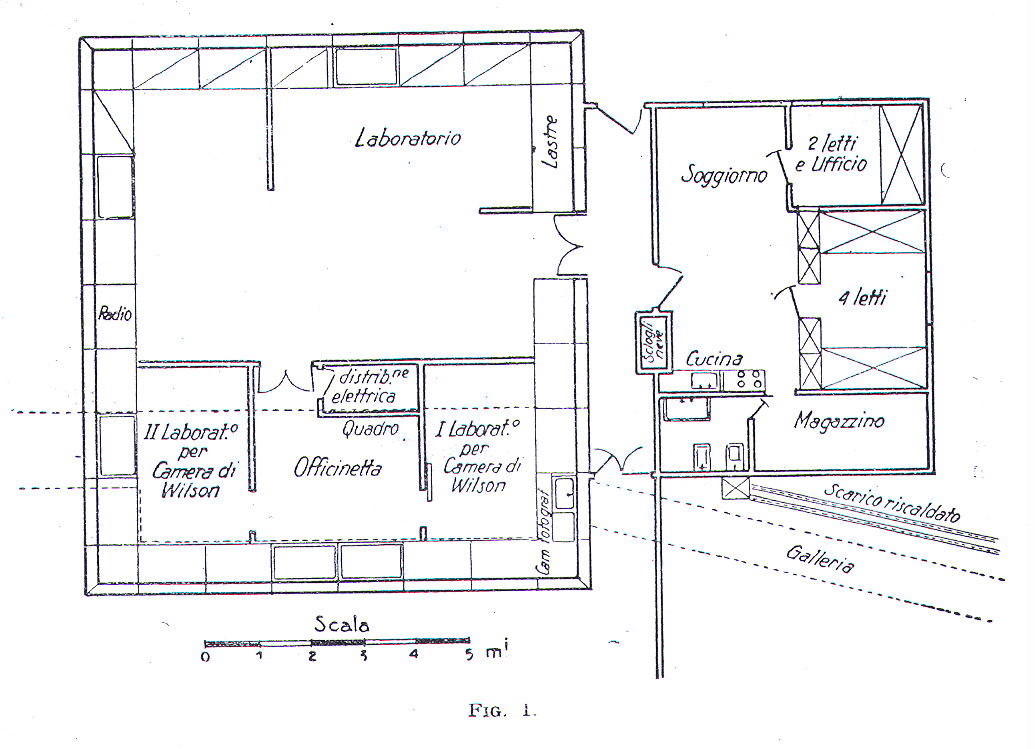}
  \caption{\label{fig:12bisB}The plant of the Testa Grigia Laboratory in 1950, reproduced from \cite{BernardiniPancini1950}.} 
\end{figure}
\end{subfigures}

Experiments were carried out there for about ten years by a large number of physicists, coming from the universities of Rome, Milan, Bologna 
and Turin. The funds necessary to set up the laboratory were 
raised from private Institutions, industrial Societies, and a 
large part of the funds of the Roman Research Centre were used 
as well. The point was that in such a laboratory ``per la sua 
altitudine [\dots] sarebbero state realizzabili ricerche sulle 
propriet\`{a} dei raggi cosmici, impossibili al livello del mare'' (cf. \cite{BernardiniLongoPancini1948}, p. 93).\footnote{``due to its altitude  [\dots] it was possible to achieve researches on the properties of cosmic rays, which were impossible at sea level''.} For instance, Bernardini and his collaborators studied there with photographic plates the absorption of the star-producing radiation -- i.e. the radiation producing nuclear disintegrations 
-- and its connection to the behaviour of the nucleonic component 
of cosmic rays (see in particular \cite{BernardiniCortiniLanfredini1948_1950}). Similar researches were performed at the Testa Grigia Laboratory by Italo Federico Quercia and Edoardo Amaldi with ionisation chambers \cite{Buschmann_et_al1950, Amaldi_et_al1950a}. Moreover, the mountain laboratory was used to study extensive air showers \cite{Amaldi_et_al1950b}.

The group of physicists who were working on cosmic rays in Milan, 
i.e. Giorgio Salvini and his collaborators, also worked at the 
Testa Grigia Laboratory in 1948-1950 to examine with a cloud 
chamber the nuclear bursts in cosmic rays \cite{LovatiMuraSalviniTagliafetti1949a, LovatiMuraSalviniTagliafetti1949b, LovatiMuraSalviniTagliafetti1949c, LovatiMuraSalviniTagliafetti1950a, LovatiMuraSalviniTagliafetti1950b}. They examined in particular the electromagnetic component arising from nuclear explosions and they also tried to provide ``material for the construction of theories on the production of mesons'' (cf. \cite{LovatiMuraSalviniTagliafetti1950b}, p. 949). In fact, one of the open questions at that time was to understand whether, and in what conditions, a nucleon-nucleon interaction produced several mesons (multiple production) or only one meson ( single production). According to the theory proposed by Walter Heitler and Lajos J\'anossy \cite{Janossy1943, Heitler1949, HeitlerJanossy1949}, ``the average fraction of energy lost by the incident nucleon in an elementary collision is comparatively small'' (cf. \cite{LovatiMuraSalviniTagliafetti1950b}, p. 949), so that the energy of the primary after a few collisions is comparable to the initial energy and ``an energetic primary should therefore give rise to successive nuclear explosions'' within 
a single nucleus (cf. \cite{LovatiMuraSalviniTagliafetti1950b}, p. 950). In other words, ``the cross sections for the production of mesons in collisions between heavy particles [\dots ] are so large that a heavy particle crossing an atomic 
nucleus must be expected to collide several times inside the 
same nucleus'' (cf. \cite{Janossy1943}, p. 345) or, as Heitler writes, ``when 
a fast nucleon passes through a compound nucleus, several mesons 
will be emitted at the same time'' (cf. \cite{Heitler1949}, p. 118; see also \cite{Rossi1952}, p. 449). This type of production was called ``plural production''. On the other hand, according to the theory supported by Heisenberg and others \cite{Heisenberg1936, Heisenberg1939, KlemmHeisenberg1943, Heisenberg1943, LewisOppenheimer_et_al1948, Wataghin1948, Wataghin1949, Fermi1950, Fermi1951}, ``genuine multiple processes 
may also exist. Then several mesons would be created in one elementary 
act (i.e., in a collision with one nucleon)''  (cf. \cite{Heitler1949}, p. 120; see also \cite{Rossi1952}, p. 449). 
As we will see, many experiments were carried out in those years 
in Italy and elsewhere to solve the question. As for Salvini 
and his group, at the Testa Grigia, they compared the explosions 
produced in nuclei of different size, and their results seemed 
to indicate that the ``incident nucleon interacts usually with 
more than one nucleon of the same nucleus'' (cf. \cite{LovatiMuraSalviniTagliafetti1950a}, p. 47). Their analysis of the development of the nuclear cascade led them to conclude that ``appare verosimile che i nostri risultati si inquadrino meglio nella rappresentazione di Heitler-J\'anossy'' (cf. \cite{LovatiMuraSalviniTagliafetti1950b}, p. 952).\footnote{``it seems likely that our results better fit Heitler-J\'anossy representation''.}\footnote{The discussion about plural or multiple production in nuclear interactions started in the mid thirties, when Heisenberg proposed to interpret the showers as explosions initiated by strong nuclear interactions \cite{Heisenberg1936}. While plural production seemed to fit with Quantum Electrodynamics (QED) predictions, multiple production seemed to indicate that at high energies, QED would show a breakdown. For a group of theoretical physicists, this limit of QED could indicate a way to modify the theory of electromagnetic interactions getting over the divergences that troubled it. From the late 30s to the 40s, new experimental evidences (bursts, stars, $\mu$-meson and $\pi$-meson) and the introduction of Yukawa's theory reopened the question of how to treat in a satisfying way nuclear interactions. 
This explains the experimental and theoretical work that was carried out to refine the knowledge on (plural or multiple) meson production from the end of the 40s to the early 50s. On the background, there was the discovery of the satisfying formulation of the late 40s QED -- with the elimination of the divergences through renormalization -- but there was also the difficulty to find a formulation of the nuclear forces theories which could be renormalized. As Pais wrote ``quantum electrodynamics looked increasingly successful but the status of meson field theories remained highly problematical''. It was not at all clear at that time -- and it was to be discovered in the following decades -- that ``the Yukawa-type interactions, unalterably important for low energy phenomena such as nuclear forces, are actually secondary manifestations of an underlying field theory, called quantum chromadynamics, not unlike the way Van der Waals forces are secondary consequences of electrodynamics. Furthermore, the Fermi interaction for weak processes, unalterably important for low energy phenomena such as beta-decay, is also a secondary manifestation of an underlying field theory -- whence the W and the Z'' (see \cite{Pais1986}, p. 551). The discussion and the researches on plural or multiple production constitute one of the starting points of these future developments.}

It is worth pointing out that the Italian physicists were perfectly aware 
at that time that cosmic rays, as a source of particles, offered 
limited possibilities as for the precision and the number of 
measures, and people knew that in those years, more and more 
powerful accelerators were being built in the US. Gilberto Bernardini wrote 
about this in 1948 ``i mesoni e le altre particelle di massa intermedia 
(oggi osservabili solo nel groviglio della radiazione cosmica) 
si potranno studiare e molto agevolmente, nei grandi laboratori 
americani dove tali macchine sono attualmente in costruzione. 
Tuttavia questo momento felice per la fisica delle particelle 
elementari, se non \`{e} immediatamente vicino per i laboratori 
degli S.U., \`{e} certamente lontanissimo per noi italiani che 
disponiamo di bilanci irrisori a favore della ricerca scientifica 
(paragonabili solo a quelli dei paesi meno progrediti del mondo)'' (cf. \cite{BernardiniLongoPancini1948}, p. 92).\footnote{``the mesons and the other particles of intermediate mass (today only observable within the twine of cosmic radiation) will be very easily studied  in the large American laboratories, where such devices are currently under construction. However this bright moment for elementary particles physics, which is not immediately close for the US laboratories, is certainly very remote for us in Italy, as we have ridiculous budgets for scientific research (only comparable to those of the less developed countries of the world)''.}

As a matter of fact, after the war, most Italian universities focused their  researches on cosmic rays, as these were the cheapest cost means to 
study particle physics. Padua, in particular, together with Rome 
and Milan, became in those years one of the most active centers 
working on cosmic ray researches. The Institute of Physics in 
Padua -- which had been planned and built by Bruno Rossi before 
the war and was at that time particularly well equipped -- had 
not been damaged and it was an important resource for the renaissance 
of physics in this university. Let us mention for instance that 
a large 7 tons electromagnet -- that Rossi had designed to deviate 
charged particles -- was ready by 1937 (fig. 14).\footnote{This device could 
generate a very intense magnetic field of about 13000 gauss with 
a gap of 15 cm between pole pieces of 28 cm of diameter \cite{Someda1937, Drigo1939}.} Rossi had also started the construction of a 1 million volt accelerator, which would have been the first accelerator of this kind in Italy 
together with the one in Rome at the Istituto Superiore di Sanit\`a; in November 1938, the setting up of the accelerator was already quite ahead but it was stopped because of the war (fig. 15).\footnote{The 1 million volt generator was 
already working. The construction of the vacuum tube, where the 
ions were to be accelerated, of the ions source, and of the voltage 
measurement devices was under way \cite{Drigo1939}.} Moreover, Rossi had planned and had made a Wilson cloud chamber to detect and photograph 
the particles' tracks. He had not been left time enough to use 
any of these instruments before his forced departure, but his successors 
were to do so after the war.\footnote{The accelerator was not to be 
really used but it was to be useful as well, as it was the starting 
point which led to the creation of the CNR Research Centre in Padua.} 

\begin{figure}

\centering

\includegraphics[width=\textwidth]{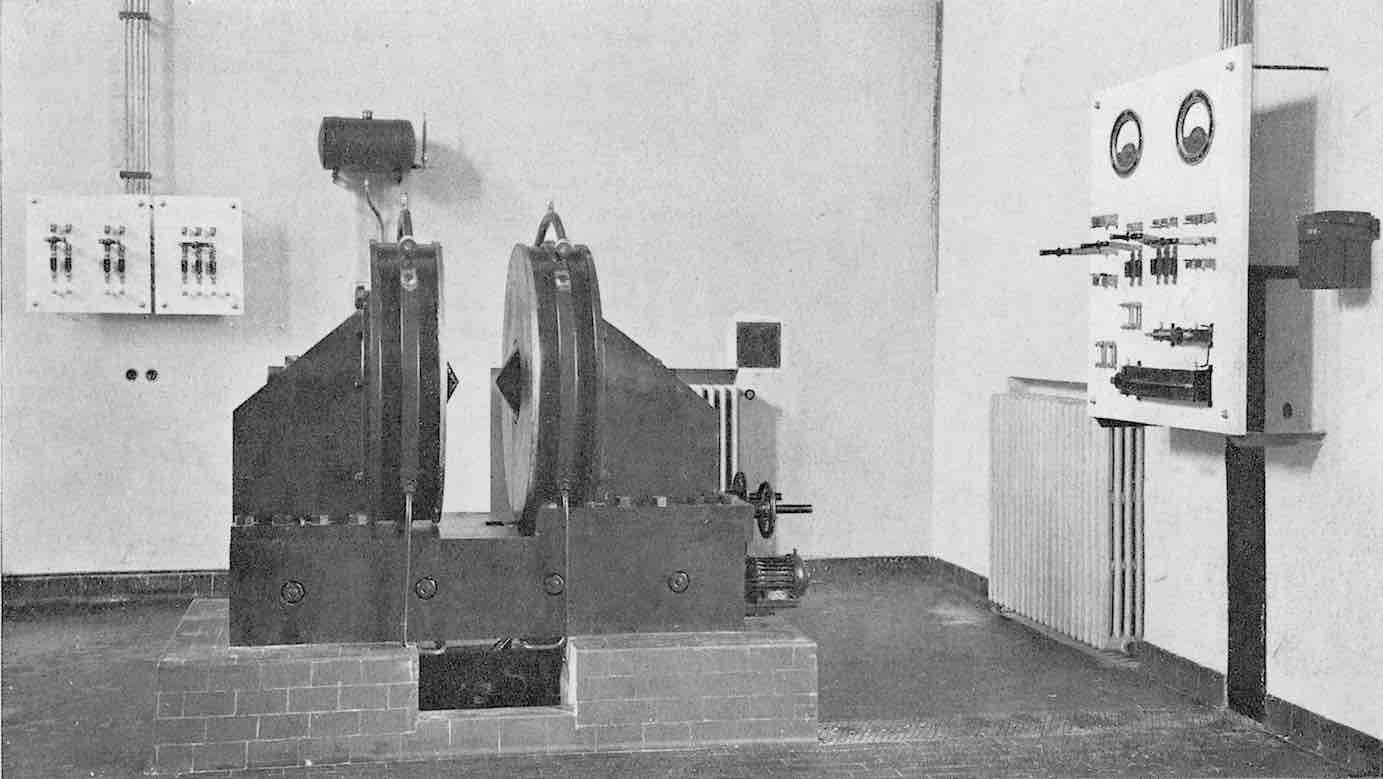}

\caption{The electro-magnet that Rossi had made as part of the equipment of the new Padua Physics Institute in 1937 (reproduced from \cite{Drigo1939}, p. 62).}\label{fig:13}

\end{figure}

\begin{figure}

\centering

\includegraphics[width=\textwidth]{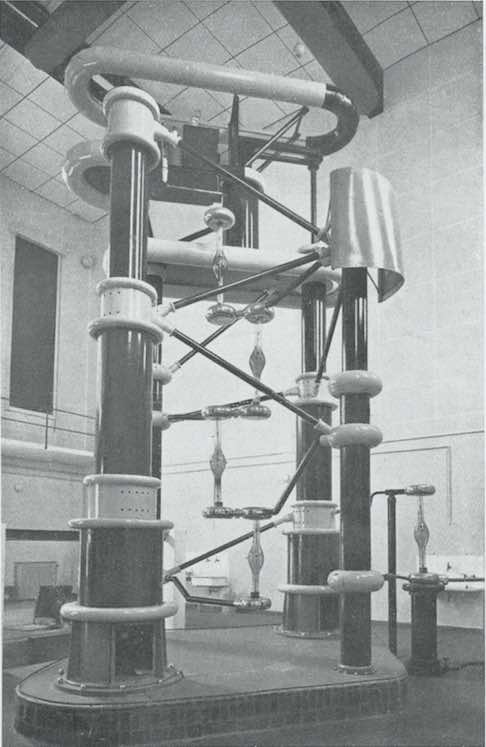}

\caption{The high voltage apparatus for the $10^6$ volt accelerator planned by Rossi in Padua, 1937 (reproduced from \cite{Drigo1939}, p. 61).}\label{fig:}

\end{figure}

In fact, Antonio Rostagni, Rossi's successor , decided at first, 
at the end of the war, to continue the construction of the 1 
million volts accelerator (fig. 16) and, in order to carry out the project, 
he achieved the creation in 1947 of a ``Centro per lo studio degli 
ioni veloci'', similar to the one that had been created in Rome 
in 1945 (see \cite{AttiCNR1947, Rostagni1949}).  

\begin{figure}

\centering

\includegraphics[width=\textwidth]{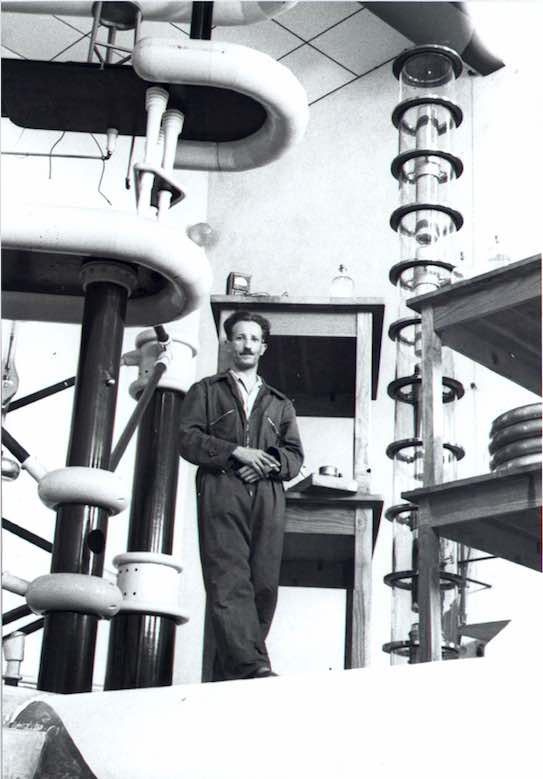}

\caption{Michelangelo Merlin working on the construction of the vacuum tube of Rossi's accelerator in the late 1940s: these attempts to make the accelerator work were unsuccessful (Dallaporta's private photographic collection - Museum of the History of Physics, University of Padua).}\label{fig:15}

\end{figure}

This Research Centre was financed by the National Research Council (CNR), exactly like the one in Rome. However, 
some technical problems and the lack of funds forced Rostagni 
to give up by 1948 the construction of the accelerator, and Padua, 
just like Rome, then decided to focus most of its activity and 
the resources that the Centro had at its disposal on cosmic ray 
researches. Different experimental techniques were used for this 
purpose. A group of physicists for instance, guided by Piero 
Bassi, started developing and using electronic instruments \cite{BassiLoria1948, BassiBeretta1949, BassiFilosofoPrinzi1950}. They studied the showers generated by the penetrating component and thoroughly examined the positive excess of the mesonic component by different means and in various conditions \cite{BassiLoria1949a, BassiLoria1949b, BassiClementel_al1949a, BassiClementel_al1949b, BassiFilosofo_al1951}. As we have already pointed out, this was regarded as a crucial question in order to understand 
the way mesons were produced by the nucleonic component. In particular, 
scientists needed more data about the value of the positive excess 
at different energies. Bassi and his collaborators also studied 
the photons within the extensive air showers \cite{BassiBianchiManduchi1951, BassiBianchiManduchi1952}. They worked both at sea level and at higher altitudes. 

In the meantime, Rostagni set up contacts with foreign laboratories: 
physicists from different European Universities were invited 
to lecture in Padua, and Paduan physicists started spending periods 
of time abroad. Arturo Loria for instance was in Manchester from 
1948 to 1951 with Patrick Blackett's group to learn more about 
the new cloud chambers techniques. Several physicists went to 
Bristol -- Marcello Ceccarelli in 1949 for instance -- and others 
went to Brussels -- Michelangelo Merlin in 1949 and in 1951 -- 
to work respectively with Cecil Powell and Giuseppe Occhialini's 
groups.\footnote{Of course, similar international 
contacts were set up in those years by the Rome physicists as 
well. Carlo Franzinetti, for instance, was also in Bristol from 
1947 to 1950 (see \cite{Amaldi1979}, p. 202-7).} The knowledge thus acquired by the Italian physicists was crucial.\footnote{For a general survey 
of the scientific activity of Padua Research Center and a complete 
bibliography of the papers produced in Padua in those years, 
see \cite{Rostagni1949, Rostagni1950, Rostagni1951, Rostagni1952}.} 

Merlin for instance, on coming back to Padua, set up in 1950 
the apparatus for the development of nuclear emulsions according 
to Dilworth-Occhialini-Payne ``Temperature Development'' method, 
and he established permanent contacts with the producing firms, 
in order to get better and better plates (fig. 17).  

\begin{figure}

\centering

\includegraphics[width=\textwidth]{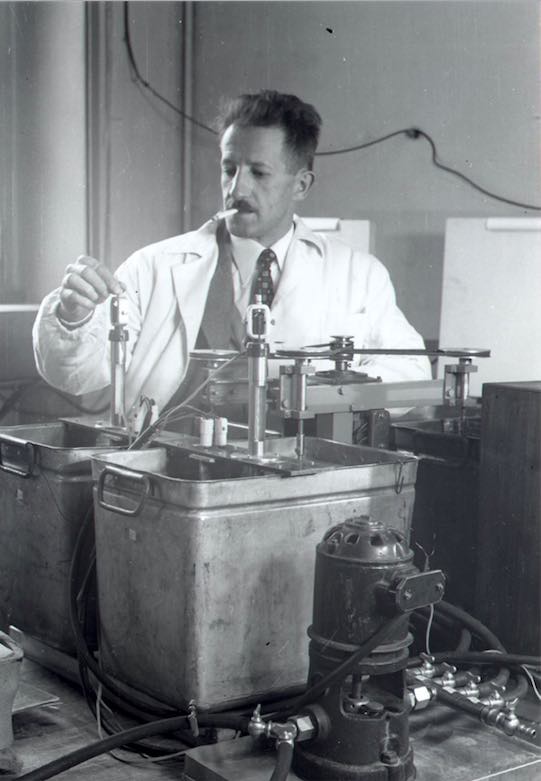}

\caption{Merlin at work in Padua with the apparatus for the development of nuclear emulsion plates set up in 1950 (Dallaporta's private photographic collection - Museum of the History of Physics, University of Padua).}\label{fig:16}

\end{figure}

The group he guided mainly worked 
on the creation of nuclear disintegration stars and on the identification 
of cosmic-ray charged particles according to the magnetic deflection 
in sandwiches of emulsion plates. Photographic plates were thus 
exposed to cosmic rays at the Institute of Physics in Padua,\footnote{See 
for instance \cite{Goldschmidt-ClermontMerlin1950}. 
These authors used the same method to study the positive excess 
of $\mu$-mesons in 1952 \cite{Merlin_al1952}.} 
on Monte Rosa (at the Capanna Regina Margherita) at 4550 meters 
above sea level,\footnote{See for instance \cite{Malaspina_al1950, Belliboni_al1951}.} and in underground galleries in the Dolomites (cf. \cite{Rostagni1952}, pp. 912-3). It is worth underlining that Rossi's 7 tons electromagnet was crucial in many of the experiments that were carried out 
in those years. As for the scanning of the plates, teams of semi-skilled persons -- called in Italy ``osservatori'' or ``lastristi'' -- were hired. They were taught to use the microscope, to recognize the events and to perform measurements. Their work, which required hours of patient observation, was of course precious and could significantly improve the scientific results of a laboratory. In Padua, there were in those years about 12 to 15 ``osservatori''; many of  them later obtained a diploma and became specialized technicians within the Institute of Physics. The knowledge acquired at that time on photographic 
plates and the contacts set up with Bristol and Bruxelles were 
to bring Padua, as we will see, to play a leading role in the 
European cosmic-ray collaborations of the early `50s. In those 
years, collaborations were also set up with other Italian Universities, 
in particular, as for photographic plates researches, with Bologna, 
Genoa (with Alberto Bonetti and his group), Milan and Pavia. Riccardo Levi 
Setti, for instance, who was working in Pavia, collaborated to 
the analysis of some of the plates exposed at 4550 meters on 
Monte Rosa \cite{LeviSettiMerlin1951}.

As for the theoretical researches on cosmic rays -- which 
were carried out by Nicol\`o Dallaporta, Ezio Clementel and others 
--, a particularly interesting contribution was given 
in Padua, in 1948-1949, by Gianni Puppi, who proposed to regard 
the $\mu$ meson as a Fermi-Dirac particle with half integer spin \cite{Puppi1948, Puppi1949}. With this hypothesis and by supposing the $\pi$-meson as a Bose particle, he obtained results which were consistent 
with the experimental data. In particular, he wrote that ``existence 
is found of a `Fermi' interaction between Fermi-Dirac particles 
(nucleons-`$\mu$' mesons-electrons) which involves the same interaction 
constant'' (cf. \cite{Puppi1949}, p.199). As Dallaporta pointed out in 1988 at the International Conference held in Rome, ``soon several data yielded definite 
support to this brilliant intuition, which has been the starting 
point towards the concept of the universality of the weak force'' (cf. \cite{Dallaporta1989}, p. 538).\footnote{It is worth noting that in the same years 1948-49 other physicists arrived independently to similar conclusions concerning the universality of the weak force (see \cite{Klein1948, TiomnoWheeler1949, LeeRosenbluthYang1949}).}  As a matter of fact, Peyrou had underlined at the 1982 \textit{Paris Colloque on the History of Particle Physics}, that Bruno Pontecorvo had suggested by June 1947 (cf. \cite{Pontecorvo 1947}) that the mesotron could have spin $1/2$ and was absorbed with the emission of a neutrino (cf. \cite{Peyrou1982}, p. 24), but Pontecorvo did not carry out any calculation on the question (in fact, almost no experimental data were available at that time, as the $\pi$-meson had not even been discovered yet).

In 1950, within the activity of the Padua Research Centre, Rostagni 
also promoted the construction, at Pian di Fedaia, at an altitude 
of 2000 meters in the Dolomites, of a laboratory for the study 
of cosmic rays (see \cite{Rostagni1951, RostagniEnElet1951} and figs. 18a, b and 19a, b),where Bassi and his collaborators carried out some of their observations by means of electronic instruments. 
\begin{subfigures}
\begin{figure}
  \includegraphics[width=\textwidth]{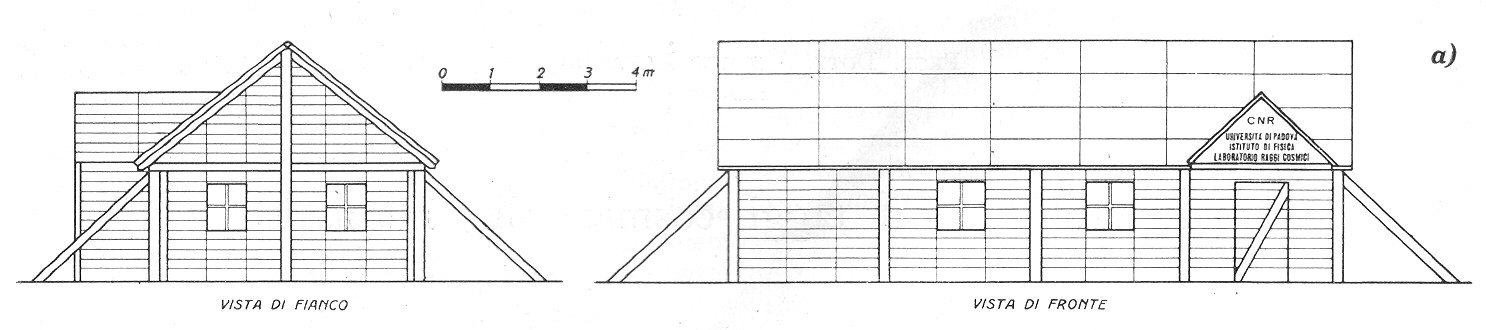}
  \caption{\label{fig:17a}Side and front view of the Fedaia Laboratory reproduced from \cite{RostagniEnElet1951}, p. 212.} 
\end{figure}
\begin{figure}
 \includegraphics[width=\textwidth]{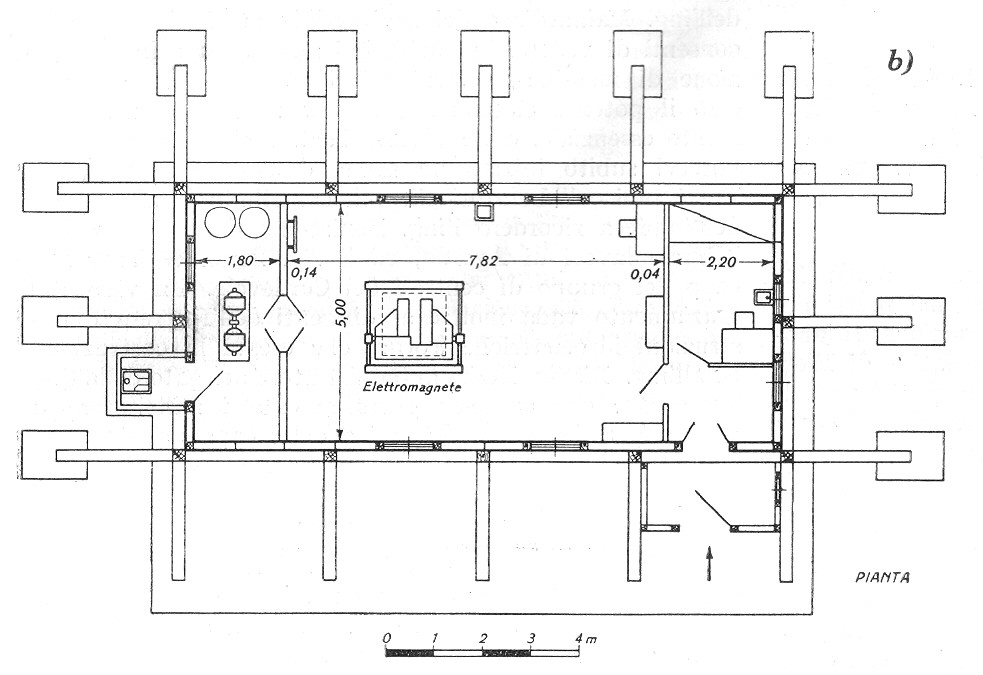}
  \caption{\label{fig:17b}Plan of the Fedaia Laboratory reproduced from \cite{RostagniEnElet1951}, p. 212.} 
\end{figure}
\end{subfigures}
\begin{subfigures}
\begin{figure}
  \includegraphics[width=\textwidth]{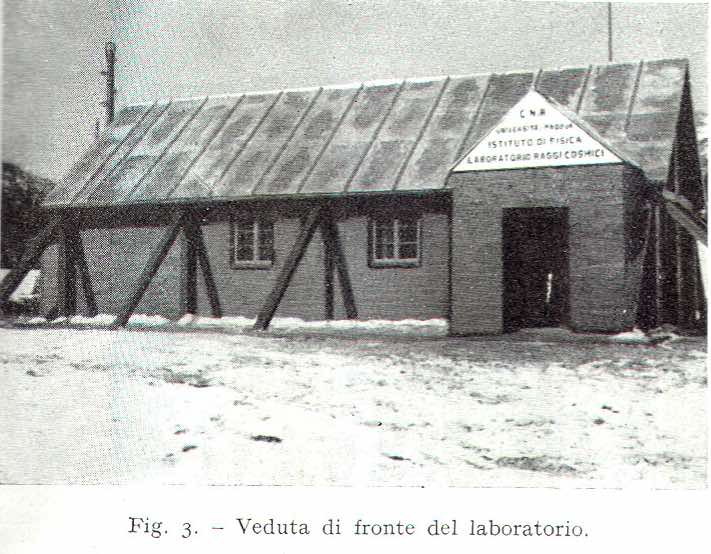}
  \caption{\label{fig:18a}Front view photo of the Fedaia Laboratory reproduced from \cite{RostagniEnElet1951}, p. 213.} 
\end{figure}
\begin{figure}
 \includegraphics[width=\textwidth]{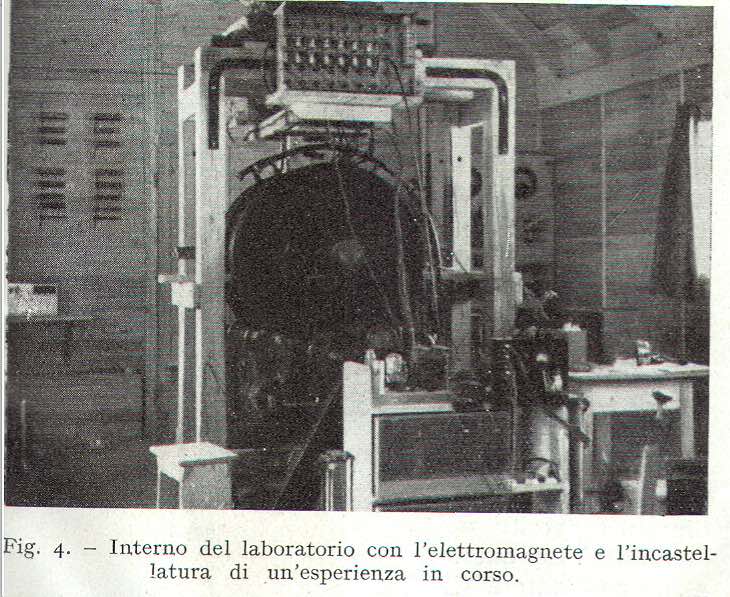}
  \caption{\label{fig:18b}Inside view photo of the Fedaia Laboratory reproduced from \cite{RostagniEnElet1951}, p. 213.} 
\end{figure}
\end{subfigures}
Another group of physicists from Padua - Marcello Cresti, Arturo Loria and Guido Zago - started working there with cloud chambers. In particular, they mounted and used at the Fedaia Laboratory the cloud chamber left by Rossi \cite{CrestiLoriaZago1953a} (see figs. 20a and b). Moreover, Rossi's 7 tons 
electromagnet was brought there as well (see figs. 18b and 19b). Cresti and his collaborators examined for instance the angular distribution of particles within extensive air showers \cite{CrestiLoriaZago1953b}. Some other experiments 
were performed in collaboration with Martin Deutschmann, one of Heisenberg's collaborators from G\"{o}ttingen Max-Planck-Institut \cite{Deutschmann_al1956}. Deutschmann was particularly interested 
in the study of the coefficient of inelasticity in collisions 
between pions and lead nuclei. Such experiments were crucial 
to verify Heisenberg's theory of the multiple production of mesons 
in nucleon-nucleon collisions, which suggested a slow variation 
of the inelasticity mean value with the primary energy. Deutschman 
brought to the Fedaia laboratory a multiplate cloud chamber from 
G\"{o}ttingen, which was used together with Rossi's chamber immersed 
in a magnetic field. The two chambers were used in pair, one 
above the other \cite{Cresti_al1956} (fig. 21). 
\begin{subfigures}
\begin{figure}
  \includegraphics[width=\textwidth]{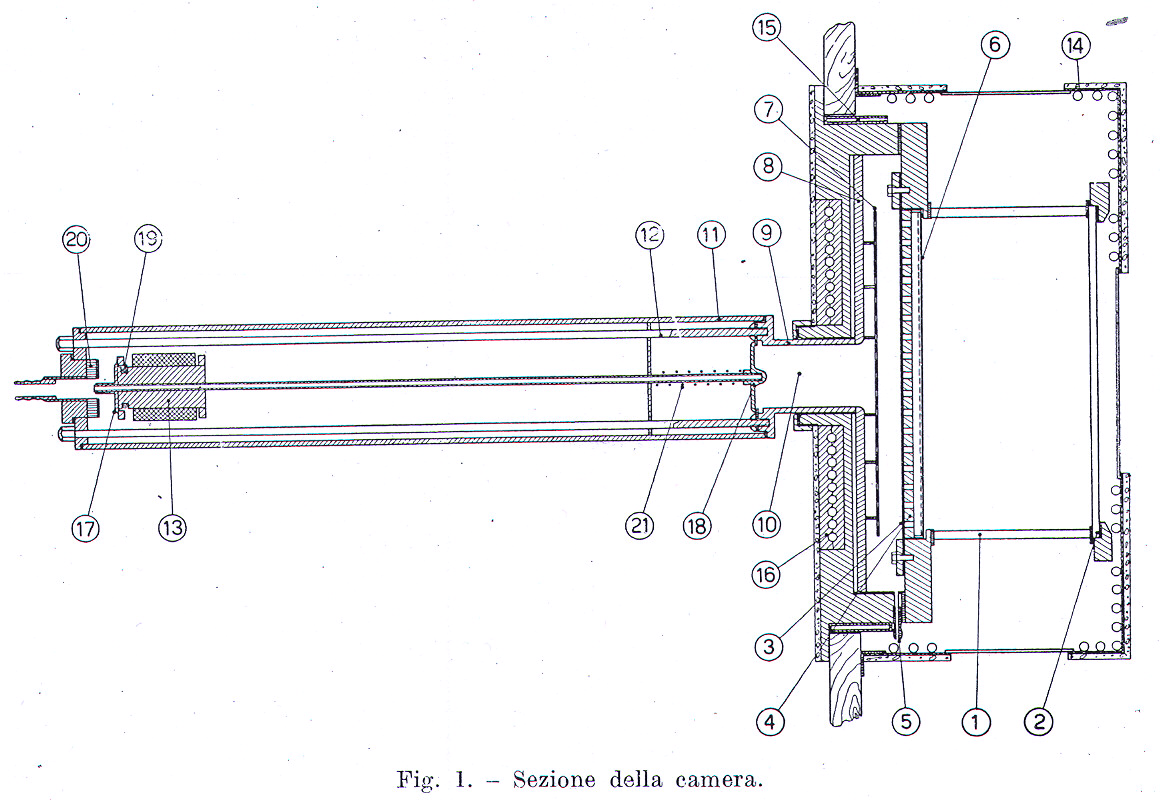}
  \caption{\label{fig:19a}Drawing of the section of Rossi's cloud chamber used at Fedaia Laboratory (reproduced from  \cite{CrestiLoriaZago1953a}, p. 844).} 
\end{figure}
\begin{figure}
 \includegraphics[width=\textwidth]{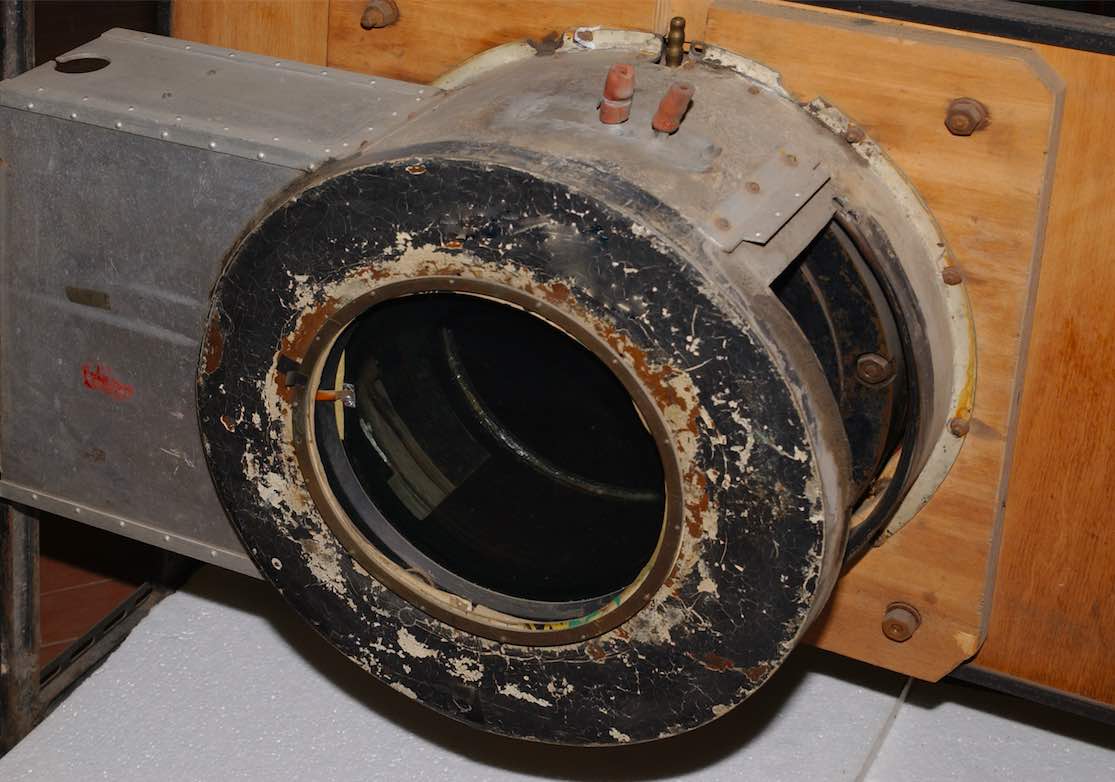}
  \caption{\label{fig:19b}A nowadays photograph of Rossi's cloud chamber used at Fedaia Laboratory (Museum of the History of Physics, University of Padua).} 
\end{figure}
\end{subfigures}
\begin{figure}

\centering

\includegraphics[width=\textwidth]{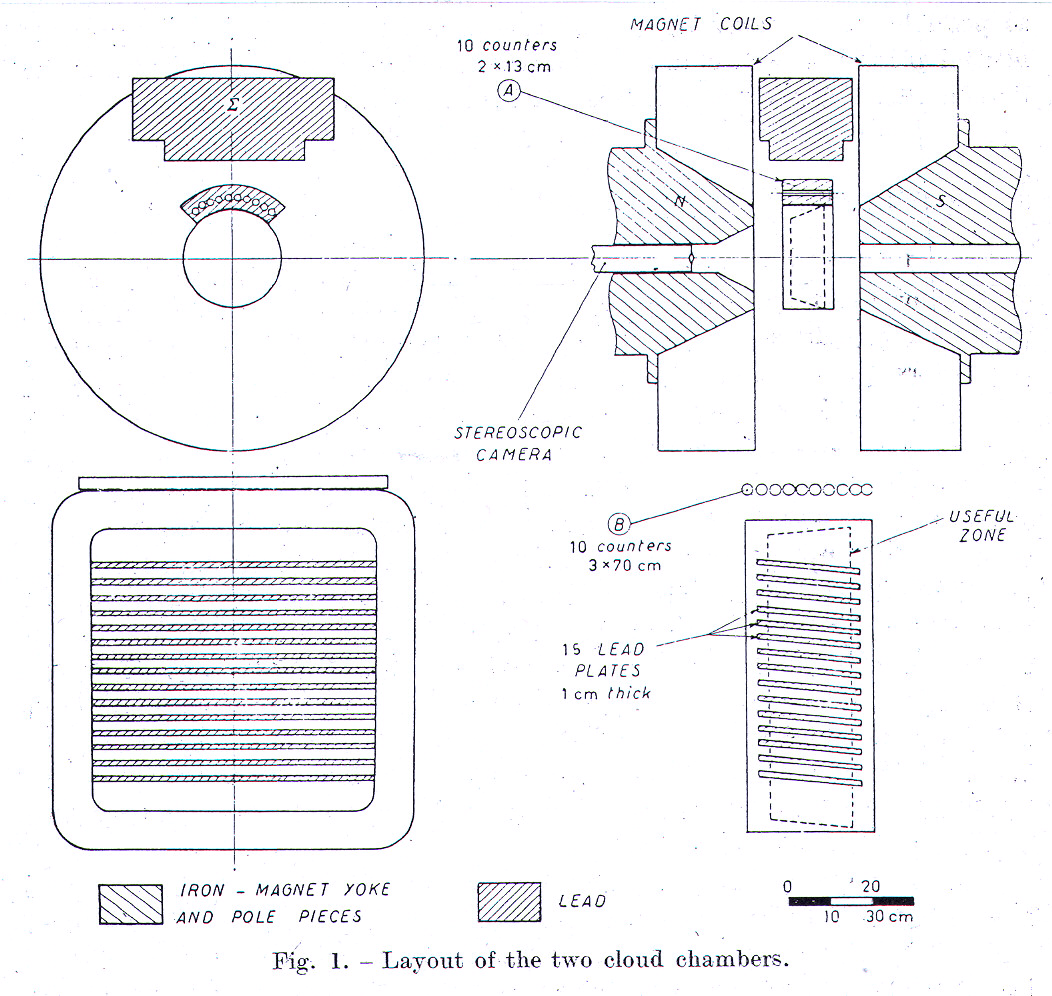}

\caption{Layout of the two cloud chambers used in pair, one above the other, at Fedaia Laboratory (reproduced from  \cite{Cresti_al1956}, p.749).}\label{fig:20}

\end{figure}
A group of physicists from London Imperial College and Edinburgh came 
to the laboratory too, in 1954, and they also brought their own 
cloud chamber, a high pressure one \cite{Burhop1955, Massey_al1956}.
As a matter of fact, only few relevant scientific results were 
achieved at the Fedaia Laboratory in the years 1951-55: as Cresti 
told us, it was too late, both because from 1946-47, other chambers 
had been made (which had been started when Rossi was finishing 
his), and because the altitude of the laboratory, about 2000 
meters, was no longer suitable (the other chambers were in use 
at about 3000 meters or more). Working at the Fedaia laboratory was nonetheless an excellent ``training'' for the young physicists who spent there 
several years - Guerriero and Cresti confirmed it -, as they 
acquired a good technical and scientific knowledge. Cresti for 
instance, after leaving the Fedaia laboratory, spent a couple 
of years in Berkeley, where he brilliantly collaborated within 
Luis Alvarez' bubble chamber group (on his second year in Berkeley, 
he officially became a member of the laboratory staff). This brought him to give interesting contributions on his coming back in Europe, in the late 1950s. As soon as he got back to Padua, for instance, he started the construction of “Franckenstein” -- a semi-automatic device for the analysis of the output of bubble chambers --, which was to be the first one in Europe, and he built an electrostatic beam separator, which was to be widely used in CERN for several years.

Padua thus started playing quite an important role at a national 
level as for cosmic ray researches. In the meantime, the group of young physicists working in Milan was quite active as 
well. Salvini and his collaborators worked in particular on the 
penetrating component of extensive air showers and they studied 
the local production of penetrating particles by extensive showers (see for instance \cite{MuraSalviniTagliaferri1947a, MuraSalviniTagliaferri1947b, SalviniTagliaferri1947, MuraSalviniTagliaferri1947c}). In 1949 Salvini 
left for Princeton, but his collaborators in Milan went on working 
on cosmic rays and, as we have said, they were soon to be joined 
by Giuseppe Occhialini, Alberto Bonetti and Livio Scarsi. 
As for the theoretical work on cosmic rays, it was mainly carried 
out in Milan by Carlo Salvetti and Antonio Borsellino. In 
Turin too, at that time, cosmic ray researches -- mainly theoretical 
-- were actively performed:  Gleb Wataghin,\footnote{Gleb Wataghin was one of the major figures of Turin physics. Born in Ukraine in 1899, he was forced to leave because of the Bolshevik Revolution and arrived in Turin as a refugee. He graduated in Physics and Mathematics and became Italian citizen in 1929. His research activity brought him to get in touch with the major physicists of those years, like Pauli, Heisenberg, Dirac,  
Gamow, Yukawa, Schr\"odinger, Bohr, Compton, Bogolubov. In 1934, partly because of the political situation, he accepted a chair at the University of S\~ao Paulo in Brazil, where he created a prestigious physics school of international level. Cesare Lattes graduated with him in 1943 and Giuseppe Occhialini worked with him from 1937 to 1942. At the end of 1948, he was assigned the Chair of Experimental physics in Turin, where he guided, together with Romolo Deaglio and Mario Verde, the rebirth of Turin physics school.} Marcello Cini, Sergio Fubini, Luigi Arialdo Radicati di Brozolo and several others gave interesting contributions (see for istance \cite{CiniWataghin1950, CiniRadicati1951, Fubini1951}). It is worth pointing out that another CNR research center, the ``Centro sperimentale e teorico di fisica nucleare'', similar to the ones founded in Rome and Padua but focused on theoretical more than on experimental nuclear physics, was created in Turin in 1951 \cite{DecretoPresCNR1951a}.
At the same time, in order to achieve a closer coordination between 
the three CNR Centers of Rome, Padua and Turin, the so called 
``Istituto Nazionale di Fisica Nucleare'' (INFN) was founded in 
August 1951 \cite{DecretoPresCNR1951b}. It was reorganized 
by July 1952: the three research centers became ``sections'' of 
the Institute and a fourth section was created in Milan \cite{DecretoPresCNR1952}. Several other universities were to join INFN in 
the following years. As Dallaporta underlined in 1988, the period 
1951-52 may thus be considered as a turning point for ``the whole 
settlement of physics in Italy. [...] The first achievement of 
this national organization was creating much more contacts and 
links between the active Italian research centers [...] moreover, 
the increase in financial support allowed a proportional increase 
in the number of research workers and technicians involved in 
the experiments [...]; finally more systematic and generally 
planned experiments extending if necessary on long periods could 
now also be undertaken'' (cf. \cite{Dallaporta1989}, pp. 538-9).

Before closing this section, we would like to point 
out that we have focused as of now on ordinary particle physics 
but, by the beginning of the 1950s, as $\pi$ mesons started to be 
produced by accelerators, the contribution of cosmic rays to 
ordinary particle physics came to an end. However, a new chapter 
in the history of particle physics had just opened, as new unstable particles, which were later to be called ``strange'' particles, had been discovered in 1947. As Charles Peyrou pointed out in 1982, this was to be ``possibly the most important contribution of cosmic rays to particle physics'' (cf. \cite{Peyrou1982}, p. 27).

\subsection{Strange particles and the huge 1950s European Collaborations on Cosmic-ray researches: the role played by Italian universities}

In 1947, soon after the discovery of $\pi$-mesons, George D. 
Rochester and Clifford C. Butler in Manchester demonstrated in 
a cloud chamber the existence of new unstable particles, with 
masses ranging from 700 to 1000 m$_{e}$ (where m$_{e}$ is the mass of the electron) \cite{RochesterButler1947}.
The new particles were called ``V particles'', because the observed events appeared in the shape of a fork. Nuclear emulsions and cloud chambers 
soon revealed many new evidences for the existence of other heavy 
unstable particles, which decayed in various different ways and 
were thus regarded at that time as several different new particles. 
There were for instance, at the beginning of 1953, V-particles, 
$\tau$-mesons, S-particles, $\kappa$-mesons, $\chi$-mesons, 
K-particles, etc.\footnote{See for instance the paper by C.F. Powell \cite{Powell1954}. This paper introduced the meeting ``A discussion on V-particles and heavy mesons'',  held 29 January 1953 at the Royal Society of London (the meeting proceedings are  published in \cite{Proceedings1953}). ``It was primarily -- as Richard Henry Dalitz writes later -- a U.K. meeting, most of the papers being from Bristol, London and Manchester, but there were reports from Paris (Ecole Polytechnique), Milan, Padua and Rome, while Butler presented a report on M.I.T. work on S particles'' (cf. \cite{Dalitz1982}, p. 196).} The interest for these new particles started growing very quickly so that, 
for instance, the 1953 Bagn\`{e}res-de Bigorre International Conference 
on cosmic rays was almost entirely devoted to the new unstable 
particles.  
It is far beyond the scope of this paper to present a complete 
survey of the history of the strange particles discovery and 
studies, which is particularly full of interconnections, of confusions 
and controversies.\footnote{Many details on this subject can be found for instance in: \cite{Peyrou1982}, pp. 28-66; \cite{Peyrou1989}; \cite{Pais1986}, in particular chapter 20; \cite{Hoddeson1989}, chapters V and VII; \cite{Milla2002}.} We will only focus here on the Italian contributions in this field, which, except for 
a few early interesting papers,\footnote{For instance, in 1949, the 
Bristol group of physicists observed in a photographic emulsion 
an event which they interpreted as due to the decay at rest of 
a particle of about 1000 m$_{e}$ into three charged mesons, probably 
$\pi$-mesons \cite{Brown_al1949}. This view was confirmed by a few other groups and the newly observed heavy particle was called ``$\tau$-meson''.  It is worth pointing out that the Padua group --Ceccarelli, Dallaporta, Merlin 
and Rostagni --  observed the $\tau$-meson decay into three $\pi$ mesons 
in 1952, on one of the plates that had been exposed on Monte Rosa \cite{Ceccarelli_al1952}. Another important Italian contribution was given in 1953 by the groups of Milan and Genoa, who published an event which established the existence of the heavy particle which was later to be called $\Sigma ^{+}$ \cite{Bonetti_al1953a, Bonetti_al1953b}.} 
were particularly important within the European expeditions 
organized in the early 1950s.\footnote{See also \cite{Amaldi1979}, pp. 220-3, and \cite{Belloni1989}.} 

The immediate task of the physicists in those years was to establish 
the properties of the new particles -- i.e. accurate value of 
their masses and lifetimes, their modes of production and decay 
-- and to see if other types of heavy mesons existed. Such researches 
could be carried out with Wilson cloud chambers or with nuclear 
emulsion plates. One of the most severe limitations of this latter 
technique lay in the observation of the tracks, which had to 
be carried out with a microscope: scanning, i.e. finding the 
tracks, and performing accurate measurements required a very 
long time, so that experiments were usually done over a few months 
or years.\footnote{The scale of the obtained tracks was on the order 
of a few hundreds micrometers and the measuring of the scattering 
angles, in particular, required time and a high precision.} This 
limitation, together with the fact that the new heavy particles 
were quite rare within cosmic rays, brought Powell to propose, 
at the Conference on Heavy Mesons held in Bristol in December 
1951, a collaboration of several laboratories in order to strongly 
increase the statistics concerning these events.\footnote{It is interesting to point out that these cosmic-ray expeditions were organised right in the years when CERN was created. They were the first great European physics research collaborations after the war. Such international collaborations marked the beginnings of Big Science in Europe and, from a social and political point of view, they represented a sign of the overcoming of the divisions and oppositions between people that had been generated by the war. In fact, they showed how science could play an important role to keep and enhance pacific relations and cooperations between European countries: a dream expressed by Niels Bohr and many other scientists.}

Powell's aim was to expose emulsion plates to cosmic rays at 
altitudes of about 20-30 km -- at the higher limit of the atmosphere 
--, so that the plates would be mainly exposed to primary radiation 
only, i.e. to heavy high energy nuclei.\footnote{Such heavy nuclei 
rarely penetrate below 20 km, as they are quickly removed, in 
passing through the atmosphere, by ionisation and nuclear collisions. 
Moreover, as a result of these nuclear collisions, a large number 
of low energy particles are produced in the atmosphere, giving 
rise to a background of effects which make the studying of the 
higher energy particles at lower altitudes much more difficult.} 
The nuclear plates were to be launched with specially made polyethylene 
balloons, which could fly at stable altitudes for several hours, 
and the obtained data were to be distributed amongst the various 
laboratories for microscopic scanning. Southern latitudes were 
more suitable for such experiments because, at low latitudes, 
more tracks of low energy particles would be removed, due to 
the action of the magnetic field of the earth.\footnote{At a given 
magnetic latitude only particles exceeding a given energy can 
approach the atmosphere. This ``cut-off'' energy is higher in southern 
latitudes. Actually, the cut-off is in momentum but it can be trasformed in a energy cut-off for a specific direction (see \cite{Johnson1938}, pp. 214-24). In fact, in the CERN Report describing the experiments only a generic ``cut-off energy'' with reference to different latitudes is mentioned (CERN Archives, CERN/16, ``Report on the Expedition to the Central Mediterranean for the 
Study of Cosmic Radiation'', p. 5).} Southern Italy was chosen, as it was at a sufficient 
low latitude and it offered good technical facilities and satisfying 
meteorological conditions as well. Two expeditions were thus 
organized, in 1952 and in 1953, and both were supported by CERN,\footnote{The so called ``provisional CERN'' was holding its very first meetings 
at that time, from May 1952 onwards (see \cite{Pestre1987}).} which regarded studies on cosmic rays as complementary to experiments with accelerators as, at that time, the high energy particles supplied by cosmic rays could not be artificially generated at all. 

The first of these expeditions was carried out from 18$^{th}$ May 
to 13$^{th}$ July 1952 and thirteen laboratories participated.\footnote{See the 
report on this expedition kept at CERN (CERN Archives, CERN/16, 
``Report on the Expedition to the Central Mediterranean for the 
Study of Cosmic Radiation''). It is interesting to point out that the 
introduction to this report was signed by both Powell and Rostagni.} 
The Universities involved were Bristol and Glasgow Universities, 
Imperial College of London, Padua, Milan, Turin, Genoa, Cagliari 
and Rome Universities, G\"{o}ttingen Max Planck Institut, the Ecole 
Polytechnique of Paris, and the Universities of Lund and Brussels. 
The balloons and the equipment were mainly contributed by the 
Universities of Bristol and Lund, where quite a considerable 
experience in flights of this nature had already been acquired.\footnote{2 
polyethylene balloons were constructed in Lund, 10 in Bristol 
and some neoprene balloons -- useful to test the radio and radar 
equipment and the meteorological conditions, but not suitable 
for level sustained flights -- were provided by G\"{o}ttingen.} In order to facilitate the recovery of the plates and to avoid the danger of exposing them to the heat of the sun -- as the Italian hinterland is mainly mountainous 
and some areas were quite inaccessible -- it was planned to allow 
the equipment to fall by parachute into the sea after a predetermined 
time.\footnote{The flights were of about 6-7 hours, so that the exposure 
would lead to 1 nuclear interaction per mm$^{3}$ of emulsion; such 
a density of events was regarded as the most convenient one for 
the microscopic scanning.} As it was believed that the winds 
at high altitude, above 70000 ft, were from east to west during 
the summer, two test flights were tried from Naples Airport, 
but the balloons were drifted in a south-east direction and the 
equipment fell on land. Cagliari, in Sardinia, was then chosen 
as the most suitable sea port in Southern Italy, as it is on 
a relatively small island and the equipment was thus expected 
to fall in most cases into the sea. The Italian Navy and Airforce 
played a crucial role, as they supplied the aircraft and the 
vessels which located, followed and recovered the equipment. 
They also granted the use of their airport instrumentation and 
provided part of the other instruments used to follow the flights, 
such as theodolites, radio-wind transmitters and receivers. 

The expedition was only partially a success, mainly because of 
the uncertain performance of the balloons, which did not always 
fly at a satisfying stable level. However, 1300 cm$^{3}$ of emulsion 
were successfully exposed at altitudes ranging from 23 and 27 
km and valuable material was obtained, which was analyzed during 
the following years.\footnote{See in particular several papers published 
in ``Il Nuovo Cimento'', in the years 1953 and 1954.} Moreover, 
the expedition gave precious details about the meteorological 
conditions in Sardinia, and valuable experience was acquired 
on several technical points, in particular concerning the launching 
and the construction of the balloons. A second expedition was 
then planned for the following year, on a larger scale and with 
the purpose of introducing several technical improvements.\footnote{See 
CERN Archives, CERN/GEN/11, ``Report on the Expedition to the Central 
Mediterranean for the Study of Cosmic Radiation -- Sardinian Expedition 
1953''. See also \cite{DaviesFranzinetti1954}.} In this new expedition, which took place in June-July 1953, nineteen universities were involved, eighteen 
of these from Europe and one from Australia.\footnote{The following 
Universities participated: Bern Bristol, Brussels, Catania, Copenhagen, 
Dublin, Geneva, G\"{o}ttingen (Max Planck Institut), London (Imperial 
College), Lund, Milan, Oslo, Padua, Paris (Ecole Polytechnique), 
Rome, Sidney, Turin, Trondheim, Uppsala.} The mains tasks were 
now divided among several universities and, in particular, three 
universities played a crucial role, Bristol of course, but also 
Rome and Padua, which became two of the organizing poles of the 
new expedition (see fig. 22).\footnote{See \cite{DaviesFranzinetti1954}, p. 482.}  

\begin{figure}

\centering

\includegraphics[width=\textwidth]{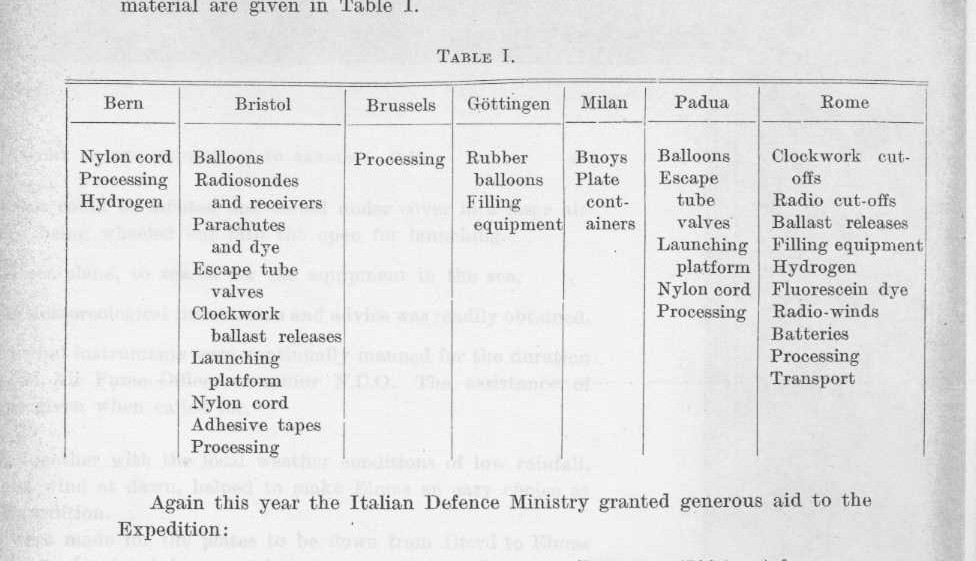}

\caption{Table describing how the tasks where distributed among the universities involved in the 1953 Sardinian expedition (reproduced from \cite{DaviesFranzinetti1954}, p. 482).}\label{fig:21}

\end{figure}

First of all, the manufacture of the balloons, which were of 
a new improved design, was carried out in Bristol and in Padua. 
Hans Heitler of Bristol University supervised the work in Padua, 
where long tables of about 30 meters were set up in the Institute 
of Physics, so that sheets of polyethylene about 2 meters large 
could be weld one with another (fig. 23). 
\begin{figure}

\centering

\includegraphics[width=\textwidth]{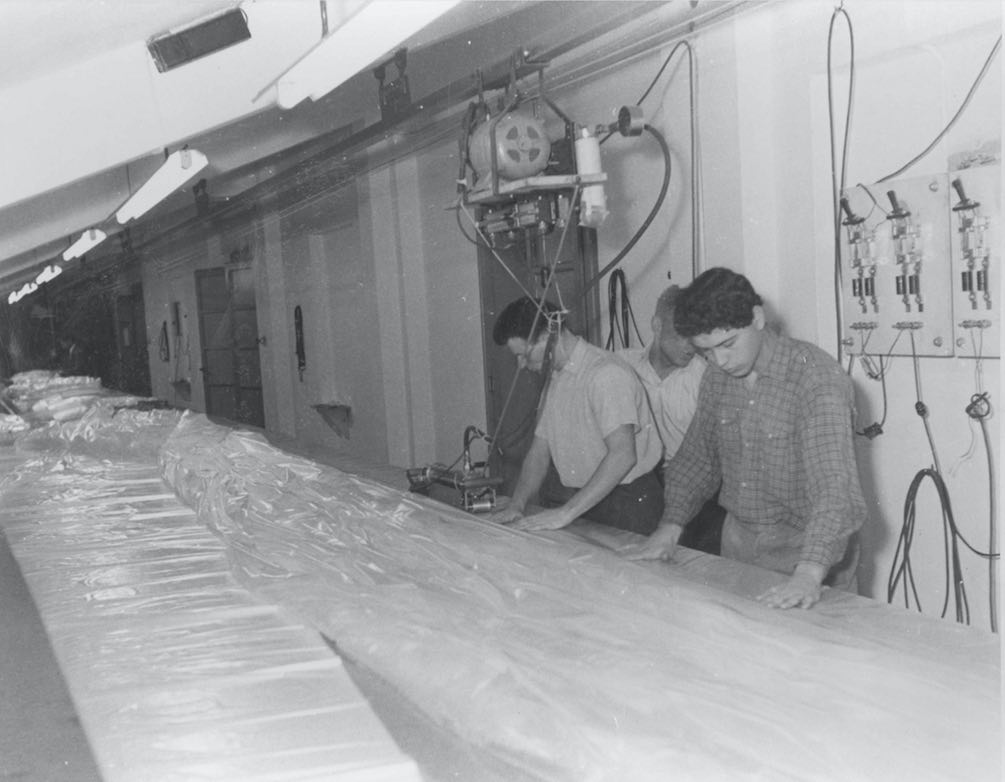}

\caption{The plant for the balloon manufacture set up at the Institute of Physics in Padua (Museum of the History of Physics, University of Padua).}\label{fig:22}

\end{figure}
A new type of machine for welding seams in the balloons 
was developed in Padua by Igino Scotoni, an engineer who became 
renowned for his brilliant capacity of developing new successful 
devices.\footnote{He was for instance to give important contributions 
to the development of bubble chambers in CERN in the 1960s.} 
This electromagnetic apparatus, ``though slower than the hot air 
device used in Bristol, was capable of sealing through several 
thicknesses of fabric'' (cf. \cite{DaviesFranzinetti1954}, p. 481, and fig. 24).
\begin{figure}

\centering

\includegraphics[width=\textwidth]{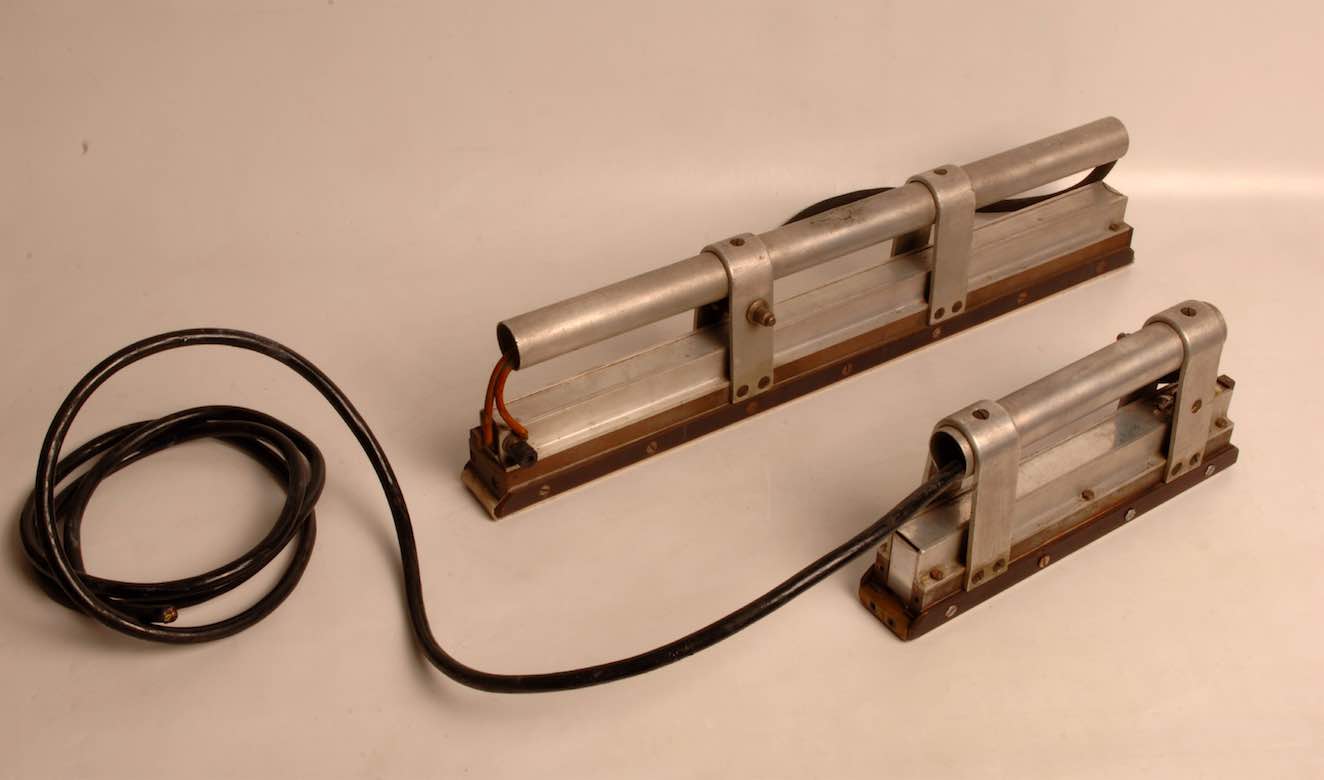}

\caption{Two of the electromagnetic devices used in Padua to weld seams in the balloons (Museum of the History of Physics, University of Padua).}\label{fig:23}

\end{figure}
Compared to the balloons used in the 1952 expedition, the new balloons were improved as for the design of the envelope, and they were also supplied with 
new escape tube valves, made in Bristol and Padua. These devices 
were to contribute to make the balloons float at the desired 
altitude, by preventing the intake of large volumes of air during 
the early part of the ascent. Twenty-three balloons were finally 
produced - thirteen of these in Padua and ten in Bristol. They 
were of different sizes, with lengths from 16 to 33 meters. They 
were to carry loads of 25-35 kg, which included the load of the 
plates (about 10 kg) and of the auxiliary equipment, i.e. clockwork 
cut-off,\footnote{It was made of an alarm clock which, after a given 
time, closed a switch connecting a battery to a resistance. The 
latter glowed and melted the nylon cord suspending the equipment.} 
parachute, sand ballast release,\footnote{Intended to prevent the 
balloon from loosing height because of small leaks and diffusion 
of hydrogen. } radio-sondes, buoys and radio-wind transmitters. 
Many of these accessories were made in Rome (see fig. 22). A few neoprene balloons supplied by G\"{o}ttingen were also used to compare the different performances. 

In order to improve the quality of the flights, two new launching 
platforms were constructed, one in Bristol and one in Padua, 
which was designed by Scotoni again (fig. 25).
\begin{figure}

\centering

\includegraphics[width=\textwidth]{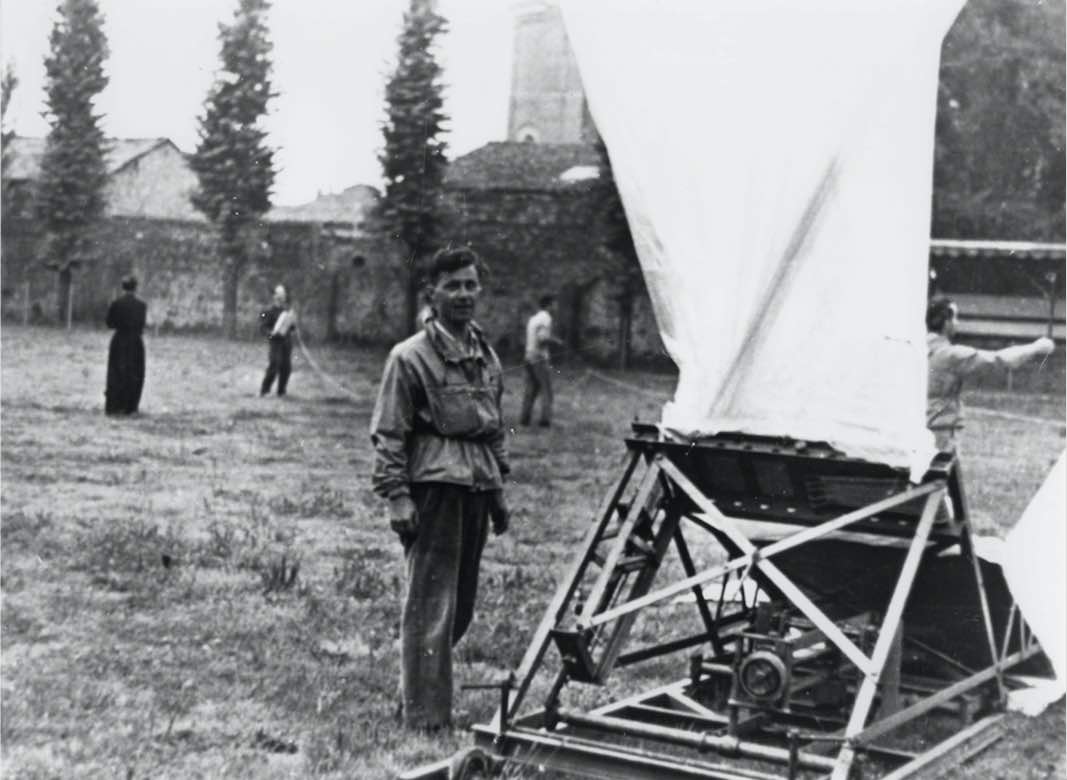}

\caption{Igino Scotoni with the launching platform he designed and built for the 1953 expedition (Museum of the History of Physics, University of Padua).}\label{fig:24}

\end{figure}
These launching platforms were intended 
to determine the right quantity of hydrogen to fill the balloon 
with, in order to predetermine the altitude of the flight. They 
were kinds of balances: the balloon and its equipment were weighed 
and the balloon was then inflated until the balance point indicated at first 
the original zero and then a free lift of about 5 kg. Twenty-three 
flights were finally carried out in Sardinia and, as in the first 
expedition, the help, support, facilities and instruments supplied 
by the Italian Navy and Air force were of crucial importance.\footnote{For 
instance, because of the high altitudes winds, the equipment 
was usually released a long way from the launching base, at distances 
of about 200-250 km, but in most cases the vessel of the Italian 
Navy successfully recovered the plates.} 

Two remaining balloons, which had not been used because of bad 
weather, were launched from Padua in the following August (fig. 26).
\begin{figure}

\centering

\includegraphics[width=\textwidth]{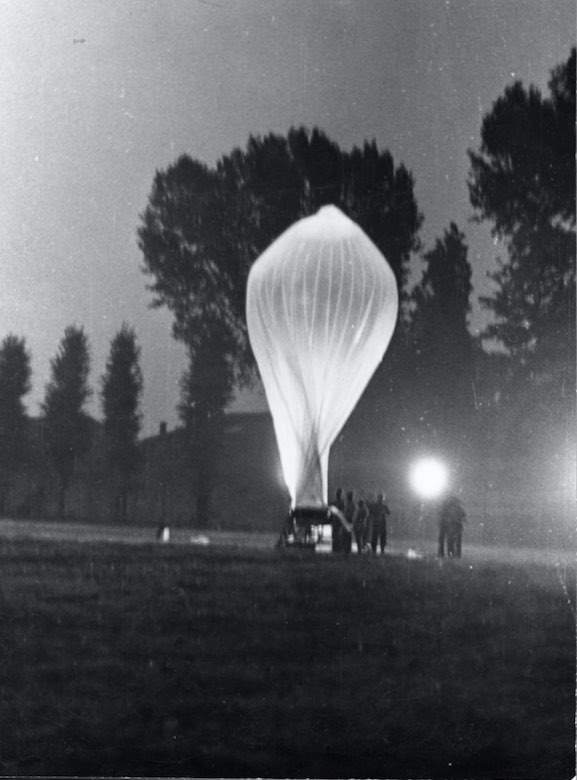}

\caption{Preparing and testing a balloon, at dawn, before one of the flights performed in Padua in August 1953 (Museum of the History of Physics, University of Padua).}\label{fig:25}

\end{figure}
In Padua, the physicists followed the flights from the ``Torre di Ezzelino'', 
one of the medieval towers of the town. They used theodolites 
lent by the University Institute of Topography and, as in Sardinia, 
they plotted on specially prepared maps the trajectories of the 
balloons. A document, fixed to the plates container, explained 
that the content was not dangerous but of important scientific 
value; people who found it were asked to bring it to the local 
authority -- ``police, mayor or pastor'' (!) -- and a sum of money 
was promised (see fig. 27). The plates of both flights 
were successfully recovered on land. 
\begin{figure}

\centering

\includegraphics[width=\textwidth]{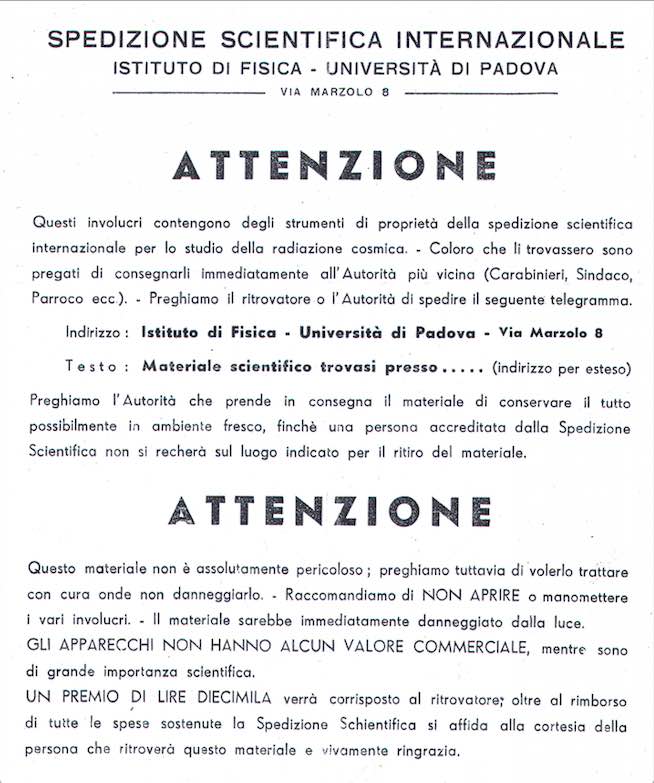}

\caption{The document fixed to the plates container explaining that the content was not dangerous but of important scientific value;  people who found it were asked to bring it to the local authority -- ``police, mayor or pastor'' -- and a sum of money was promised (Museum of the History of Physics, University of Padua).}\label{fig:26}

\end{figure}

On the whole, 22 stacks of 40 emulsions (each $15\times10\times0.06$ cm$^3$ in size) were recovered from 37 flown, so that about 9 liters 
of emulsion were exposed. The plates then had to be developed, 
and the processing, which had been mainly performed in Bristol 
in 1952, was now also carried out in Rome, Padua, Bern and Brussels. 
The plates were, as before, supplied by Ilford (Bristol), but 
it is important to point out that for this second expedition, 
the so called ``stripped emulsions'' were used, i.e. large stacks 
of emulsion layers, each of which had been stripped off its glass 
backing \cite{Powell1953}. Each layer was 
thus in direct contact with the next one, and this was an important 
improvement as, by eliminating the glass, one could follow the 
tracks of the particles from one layer to another over much larger 
distances and it was possible to carry out measurements with 
much more precision. Of course, new techniques had to be introduced 
to work with stripped emulsions. For instance, before the development 
phase, each individual emulsion was made to adhere to a specially 
treated glass plate. This was done by dipping the emulsion in 
a purposely prepared solution and placing it on the glass plate; 
the latter was then passed between the rubber lined rollers of 
an ordinary domestic mangle. The pressure between the rollers 
had to be chosen carefully in order to limit the distortions.\footnote{See 
\cite{Powell1953}, p. 221. One may wonder whether such a treatment  introduced distortions which made difficult the analysis of the plates. In fact, it did introduce distortions, but mainly on a larger scale, not relevant 
at the distances of a few microns analyzed in these cases.} 

Once processed, the plates were distributed among the members 
of the collaboration according to their financial and effective 
support. As money was needed to reduce the financial deficit 
of the expedition, some stacks were sold to the Universities 
of Trondheim, Paris (Ecole Normale) and Warsaw. As for the scanning 
of the plates, about a hundred microscopes were used throughout 
the different laboratories involved. Only the most perfected 
microscopes available commercially in those years for biological 
and medical research could be used for emulsion technique and 
then, only for the simplest operations, i.e. scanning and location 
of the events. To perform measurements, in particular scattering 
measures on the tracks, the motions were actuated by too coarse 
screws and stage graduations were not precise enough. Microscope 
manufacturers were only beginning in those years to work on this 
new domain, and a few special models, specifically 
intended for emulsions, were made by a few manufacturers -- as for 
instance Cooke, Troughton and Simms in England, Koristka in Italy 
and Leitz in Germany (cf. \cite{Goldschmidt-Clermont1953}). Such 
special models were usually developed through a close collaboration 
between the physicists and the microscope workshops. 

In the meantime, while the 1953 flights were under way, the Bagn\`{e}res-de-Bigorre Conference took place and, as Milla Baldo Ceolin points out ``it 
soon became clear that all this widespread cosmic-ray work was 
leading to a substantial consensus. Previously it had seemed 
as if a new decay mode, or perhaps a known decay mode for a new 
parent mass, was being reported almost every month. But now it 
turned out that the most frequent decay modes were quite limited 
in number, and they were associated with fairly definite mass 
values. Therefore, a coherent picture of the new particle physics 
began to emerge from many partial works; in attempt to classify 
the new particles, we established together the existence and 
the properties of many particles'' (see \cite{Milla2002}, p. 7). Soon 
after the Conference, Louis Leprince-Ringuet divided the mesons 
and heavy unstable particles into three groups: ``L-mesons (including 
$\pi$-mesons, $\mu$-mesons, any other possible lighter meson)'', ``K-mesons 
(particles with mass intermediate between those of the $\pi$-meson 
and the nucleon)'', and ``H-particles or Hyperons (particles with 
mass intermediate between those of the nucleon and the deuteron)'' (cf. \cite{Leprince-Ringuet1953}, p. 64). The K-meson sector looked particularly confused, as many different decay modes had been observed. There were the so called $\tau$ particles, the $\kappa$ particles, the $\chi$ particles and the $\theta^{0}$. Striking similarities appeared between these 
different groups of particles as for their lifetime and their mass.

The first results of the scanning operations of the 1953 expedition 
were presented at the International Conference held in Padua 
in April 1954, one year after the Bagn\`{e}res-de-Bigorre Conference\cite{CongInterPadova1954}.\footnote{Many other results obtained 
from the Sardinian stacks were published later, up to several 
years after. See for instance the paper \cite{Milla_al1955}, which summarizes the work carried out by the Padua group on the photographic plates exposed in Sardinia in 1953 , and the paper giving the results obtained by the Milan 
and Genoa group \cite{Bonetti_al1955}.} 
Thanks to the technical progresses achieved -- connected in particular 
to the use of stripped emulsions --, both the characteristics 
of the events from which the unstable particles emerged and their 
mode of decay had been determined in many cases. Moreover, it had been possible to follow many particles to the end of their range, so that much 
more precise results had been obtained concerning the masses 
of the new particles. At this conference, where data were also 
provided by cloud chamber experiments and by accelerators, several 
typical aspects of the phenomenology concerning the new particles 
were focused and, as Dallaporta wrote in 1988, the Conference 
thus brought a first ``clarification on the different types of 
decay of the K mesons, with some indication that a single type 
of particle was decaying into several competitive modes'' (see \cite{Dallaporta1989}, p. 540). However, some important uncertainties 
survived. One of these was connected to $\tau$ and $\theta$ particles. 
As Baldo Ceolin wrote in 2002, ``The key question was: Could the 
$\theta^{0}$ particle, namely the K$^{0} \rightarrow \pi^+ \pi^-$, and the 
$\tau^+$ particle (K$^{+} \rightarrow \pi^+ \pi^+ \pi^-$) be closely related? 
Both $\tau$ and $\theta$ decayed to pions only and their masses were 
quite comparable. One would have expected them -- following the Rossi argument -- to be different decay modes of the same particle. The $\pi^+ \pi^-$ state resulting from $\theta^0$ decay, since the $\pi$-meson is a pseudoscalar particle, has parity $(-1)^J$, where $J$ is the relative orbital angular momentum, so $0^+$, $1^-$, $2^+$, ...  The question was, could the $\pi^+ \pi^+ \pi^-$ state resulting from $\tau^+$ decay have the same spin and parity as the $\pi^+ \pi^-$ state resulting from the $\theta^0$?'' (cf. \cite{Milla2002}, pp. 10-1). Another question raised at the Padua Conference 
was related to the discovery, presented by the Ecole Polytechnique group, 
of another K-meson decay scheme, the so called K$_{\mu 2}$ decay \cite{Gregory_al1954, Armenteros_al1955}. Such a decay was discovered 
with cloud chambers and it was quite a surprise, as it had not 
at all been observed in emulsions up to that moment. Emulsion physicists where quite puzzled and embarassed.

All the uncertainties left and the open questions led some of 
the emulsion groups present at the Padua Conference to plan a 
new experiment, the so called ``G-stack'', or ``Giant'' stack, mostly 
prompted by the Universities of Padua, Milan and Bristol. In 
October 1954, another polyethylene balloon was thus launched 
from Novi Ligure, in Italy, and a single huge stack with 15 liters 
of emulsion (i.e. 63 kg!), with the dimensions of $37\times27\times15$ $\mbox{cm}^{3}$, was exposed to cosmic rays in the stratosphere. Such a large 
stack of emulsion was supposed to offer full containment of many 
tracks of the decay products of K-mesons. In other words, if 
the emulsion stack was large enough, any of the secondaries would 
stop in the emulsion and could thus be thoroughly studied. Concerning 
this new expedition, Powell wrote, ``I think my colleagues and I will all agree that if anybody has played a distinctive and leading part, it is Dr. {\sc Merlin}. He played a very important role in the early 
days during the discussions on the feasibility of flying a 
very large stack; throughout the expedition his enthusiasm and 
his confidence in a successful outcome were of the greatest importance; 
and finally, his drive and enthusiasm in the period of the examination 
of the plates were largely responsible for the fact that the whole 
enterprise was brought to a successful conclusion. We are all of us greatly in his debt'' (cf. Powell's ``Introduction'' to \cite{GStack1956}, p. 400-1).  Actually, the results of the G-stack experiment, presented at the Pisa International Conference on Elementary Particle in 1955, gave the definitive 
information on the K-meson decay modes \cite{GStack1955, GStack1956}. The G-Stack group -- quite a huge working group for those times, composed of 36 scientists --\footnote{The authors were from the universities of Bristol, Copenhagen, Dublin, Genoa, Milan, Bruxelles and Padua.} was of course enthusiastic and Cormac O'Ceallaigh, who was part of the group, described their preparing the results for the Pisa Conference with these words 
``The organisation of the results of the measurements and their presentation was again, largely the work of the Italians and took place at Padova. I never will forget the fever and the excitement associated with the effort. Occhialini was in ultimate charge and strode up and down the 
scene like an avenging Jehovah or thundering Jove, cursing and swearing and casting unjust aspersions on the ancestry and sexual morals of any wight unlucky enough to have blundered. He was just as quick to apologise when the dust had settled'' (cf. \cite{O'Ceallaigh1982}, p. 187). In fact, the G-Stack collaboration ``established beyond any doubt that the phenomenologically different decay modes observed up to then [...] were due to a single type of 
particle, which since then has been termed the K$^+$ meson; the 
$\theta^0$ was its neutral counterpart'' (cf. \cite{Milla2002}, p. 
13).

It is interesting to underline that at the Pisa Conference, the 
new data about the $\tau$-decay -- most of which coming from the 
G-Stack -- were also examined and this led to a clear picture 
of the so called $\tau$-$\theta$ puzzle: $\theta$ and $\tau$ had finally been 
identified as one single particle, and it was thus clear 
that two decay modes of the same particle had different parities. 
This question was to be solved a few years later. In 1956 Tsung Dao Lee and Chen Ning Yang \cite{LeeYang1956} proposed the hypothesis of parity non conservation in weak interactions, suggesting some experiments to verify the issue. During the following months, such experiments were carried out confirming Lee and Yang proposal. The parity violation can 
be regarded as ``the most astonishing and revolutionary conclusion 
of the cosmic-ray emulsion era''(cf. \cite{Milla2002}, p. 15).\footnote{It is worth pointing out that previous theoretical and experimental results contributed to this revolutionary conclusion. In particular, Dalitz had proposed in 1953-54 (see \cite{Dalitz1953_54} and \cite{Dalitz1982}) a method for the analysis of the $\tau$-meson decay process in terms of its spin-parity. This method, which became known as ``Dalitz Plot'', was independently proposed by Elio Fabri \cite{Fabri1954}, a young theoretician linked in Rome to Bruno Ferretti and Bruno Touschek.}

On the whole, the Sardinian flights and the G-stack were therefore 
a great success for cosmic-ray physicists, as the basic properties 
of the new unstable heavy particles were now known. However, 
at the Pisa Conference, the first results on the K properties 
from Berkeley Bevatron were also presented, showing that accelerators 
could now provide large statistics and accurate measurements 
not only for low energy particles but for particles of higher 
and higher energy: cosmic rays as sources of particles were soon 
to be completely superseded by accelerators, and the Sardinian 
and G-stack expeditions were to be the last important cosmic-ray 
collaborations. In Pisa, ``cosmic-ray physicists (emulsion and cloud-chamber experts) celebrated [their] final triumph'', and this happened 
``just a few years after the real beginning'' (cf. \cite{Milla2002}, p. 
13). At about the same time, the nuclear emulsions groups in 
Italy started collaborating with some of the accelerators groups 
in the US. They were provided with stacks that had to be processed 
and analyzed, and they went on carrying out researches with 
photographic emulsions for several more years, but cosmic rays 
were no longer their source of high energy particles.

It is worth underlining that by that time, Italian physics had 
reached the level of the greatest scientific centers in the world, 
and this extraordinary rebirth had taken place in a few years 
only, precisely during the ``cosmic ray period''. According to Amaldi, the 
turning point was the mid fifties, when ``gli anni della ricostruzione 
erano chiaramente terminati grazie ad un'opera collettiva non 
molto frequente nel nostro paese per ampiezza numerica, variet\`{a} 
e qualit\`{a} delle persone e durata nel tempo (circa un decennio). Le 
stesse strutture organizzative erano veramente nuove e avrebbero 
potuto servire d'esempio per altre attivit\`{a}, non solo di fisica 
pura e applicata, ma di ricerca in generale. Cominciava dunque, 
in Italia, anzi in Europa, una nuova fase, non solo per le ricerche 
relative alle particelle elementari ma in generale per tutta 
la ricerca'' (see \cite{Amaldi1979}, p. 225).\footnote{``the years of reconstruction were clearly over, thanks to a collective action quite unusual in our country as for the numerical size, the variety and quality of the persons and the length (about a decade). The very organizational structures were completely new and could be an example for other activities, non only of pure and applied physics, but of research in general. A new phase was thus starting in Italy, or rather in Europe, not only for the researches related to elementary particles but for research in general''.}

Finally, with reference to the further developments of particle physics, we would like to conclude by quoting a passage from the talk Peyrou gave in Rome in 1988: ``I would like to answer to some people who marvel why so much effort was spent by cosmic ray physicists to settle questions which were easy to answer with the accelerators. There are two answers: (a) The accelerators physicists went very fast because the cosmic rays people had completely clarified the zoology before they came; (b) it is a rule of physics and I think of science in general that knowledge is won the hard way whereas things might have been easier to find years later, but if it were not done that way science will probably come to a stop'' (cf. \cite{Peyrou1989}, p. 632-3). 

\acknowledgments
We would like to express our gratitude to the library staff of the
Padua Department of Physics, i.e. Alessandra Barbierato, Germana
Bertante and Fernando Tavazzi, and to the CERN Archivist, Mrs Anita Hollier. We are greatly indebted for wise comments and discussions to Milla Baldo Ceolin, Giancarlo Bettella, Marcello Cresti, GianAntonio Salandin, and to all the participants to the Padua Meeting ``La rinascita della fisica in Italia dal secondo dopoguerra ai primi anni 1960: protagonisti a confronto'', held in
Padua in September 2006. We would also like to warmly thank Jean-Pierre Hurni for his devoted support in and from Geneva. 

Finally, we express our gratitude to an anonymous referee who gave us the opportunity of improving the original manuscript.

\end{document}